\documentclass[aps,twocolumn,showpacs,preprintnumbers,floats]{revtex4}\def\@cite#1#2{\textsuperscript{[{#1\if@tempswa , #2\fi}]}}
\usepackage{mathrsfs}

\usepackage{xcolor}
\usepackage{longtable,lscape}
\usepackage{savesym}
\usepackage{txfonts}
\savesymbol{iint}
\savesymbol{iiint}
\savesymbol{iiiint}
\savesymbol{idotsint}
\usepackage{amssymb}
\usepackage{amsmath}
\restoresymbol{TXF}{iint}
\restoresymbol{TXF}{iiint}
\restoresymbol{TXF}{iiiint}
\restoresymbol{TXF}{idotsint}
\usepackage{indentfirst}
\usepackage{graphicx,booktabs}
\usepackage{braket}
\usepackage{multirow}
\usepackage{color}
\usepackage{subfigure}
\usepackage{float}
\usepackage{bm}
\usepackage{epsfig}
\usepackage{booktabs}
\usepackage{tikz}
\usepackage[colorlinks, citecolor=blue,anchorcolor=red,menucolor=red, linkcolor=red,filecolor=red,urlcolor=blue,frenchlinks=red]{hyperref}
\usetikzlibrary{calc} 

\begin{document}
	
	\title{Exploring the properties of low-lying excited $\Xi_c$ and $\Xi_c'$ baryons}
	\author{Yi-Heng Wang$^{1}$, Xin-Yu Chang$^{1}$, Hui-Hua Zhong$^{2}$~\footnote {E-mail: zhonghuihua@foxmail.com}, Li-Ye Xiao$^{1}$~\footnote {E-mail: lyxiao@ustb.edu.cn}}
\affiliation{ 1)Institute of Theoretic
al Physics, University of Science and Technology Beijing,
Beijing 100083, China}
\affiliation{ 2)School of Aeronautics Mechanical and Electrical Engineering, JiangxiFlight University, Nanchang 330088, China}

\begin{abstract}
Motivated by the rapidly accumulating experimental data, we systematically calculate the $\Xi_c$ and $\Xi_c'$
mass spectra for the $1S$-, $1P$-, and $1D$-wave states and investigate the strong decay properties of the $1P$- and $1D$-wave states, thereby interpreting the internal structure of the currently controversial excited states and providing useful guidance for future experimental searches. The main conclusions are summarized as follows. $\Xi_c(2790)$ and $\Xi_c(2815)$ can be well explained as $\Xi_c1P\big|J^P=1/2^-,1\big\rangle_{\lambda}$ and $\Xi_c1P\big|J^P=3/2^-,1\big\rangle_{\lambda}$, respectively. $\Xi_c(2882)$ may correspond to $\Xi_c1P\big|J^P=1/2^-,0\big\rangle_{\rho}$ or $\Xi_c1P\big|J^P=1/2^-,1\big\rangle_{\rho}$; under this interpretation, $\Xi_c\pi$ and $\Xi_c'\pi$ would be promising channels for its detection. $\Xi_c(2923)$ might be assigned to $\Xi_c'1P\big|J^P=3/2^-,1\big\rangle_{\lambda}$, but further confirmation is required. $\Xi_c(2939)$ is very likely to be $\Xi_c'1P\big|J^P=3/2^-,2\big\rangle_{\lambda}$ and has a good potential to be observed in the final states $\Xi_c\pi$ and $\Xi_c'^{*}\pi$. $\Xi_c(2965)$ can be explained as $\Xi_c'1P\big|J^P=5/2^-,2\big\rangle_{\lambda}$ and $\Xi_c\pi$ decay channel is a promising additional mode for future experimental searches. $\Xi_c(3055)$ and $\Xi_c(3080)$ can be assigned as the quark configurations $\Xi_c1D\big|J^P=3/2^+,2\big\rangle_{\lambda\lambda}$ and $\Xi_c1D\big|J^P=5/2^+,2\big\rangle_{\lambda\lambda}$, respectively. $\Xi_c(3123)$ is possibly state $\Xi_c'1D\big|J^P=1/2^+,1\big\rangle_{\lambda\lambda}$ or $\Xi_c'1D\big|J^P=5/2^+,3\big\rangle_{\lambda\lambda}$, with the latter being more favorable. Decay channels $\Lambda/\Sigma D$ are very likely to serve as a criterion for distinguishing between these two possible candidates.

\end{abstract}
	\maketitle

\section{Introduction}

The single-charm strange baryons, states with quark content $csd$ or $csu$, have a wave function that is antisymmetric(denoted $\Xi_c$ baryons) or symmetric(denoted $\Xi_c'$ baryons) under interchange of the light quark flavors. Due the large mass difference between the charm and lighter quarks, these baryons serve as an important system for studying heavy-quark symmetry, light-quark two-body correlations, and SU(3) flavor symmetry breaking.

For the $\Xi_c$ and $\Xi_c'$ baryons, three isospin doublets of ground states are expected in the quark model~\cite{Gell-Mann:1964ewy,Zweig:1964ruk,Klempt:2009pi}. These states were successively observed experimentally between 1983 and 1999 by the CERN WA62~\cite{Biagi:1983en} and CLEO~\cite{CLEO:1988yda,CLEO:1998wvk,CLEO:1996zcj,CLEO:1995amh} Collaborations, and were subsequently confirmed by a series of other experiments. Meanwhile, significant progress has also been made in the observation of excited states. In 1999, the CLEO Collaboration observed the first orbital excited state $\Xi_c(2815)$ with spin-parity $J^P=3/2^-$ in the $\Xi_c^*\pi$ decay channel~\cite{CLEO:1999msf}. Two years later, another orbital excited state $\Xi_c(2790)$ with spin-parity $J^P=1/2^-$ was reported by the same collaboration in the $\Xi_c'\pi$ decay channel~\cite{CLEO:2000ibb}. The masses and decay properties of these two orbital excited states were further measured precisely by the Belle and LHCb Collaborations~\cite{Belle:2016lhy,Belle:2020ozq,LHCb:2022vns,Belle:2008yxs,LHCb:2025mge}. In 2006, two additional states $\Xi_c(2970)$ and $\Xi_c(3080)$ were reported by the Belle Collaboration in the $\Lambda_c K \pi$ final state~\cite{Belle:2006edu}. Subsequently, these two states were successively further confirmed by later experiments~\cite{BaBar:2007zjt,Belle:2013htj,Belle:2016tai,LHCb:2025mge,Belle:2016lhy,Belle:2020tom,Belle:2008yxs}. It is worth emphasizing that the Belle Collaboration determined the spin-parity of $\Xi_c(2970)$ to be $1/2^+$ through angular distribution analysis in 2020~\cite{Belle:2020tom}, a result that serves as an anchor for determining the internal structures of the observed excited states. The $\Xi_c(2930)$ was initially observed by the BaBar experiment~\cite{BaBar:2007xtc} and later confirmed by the Belle Collaboration~\cite{Belle:2017jrt,Belle:2018yob} via the $\Lambda_c K$ mass distribution. With increased statistics, the LHCb collaboration resolved the $\Xi_c(2930)$ into two narrow states, $\Xi_c(2923)$ and $\Xi_c(2939)$~\cite{LHCb:2020iby,LHCb:2022vns,LHCb:2025mge}. Meanwhile, the LHCb collaboration reported another $\Xi_c$ resonance candidate, $\Xi_c(2965)$~\cite{LHCb:2020iby}. Although its mass is close to that of the previously observed $\Xi_c(2970)$, its width differs significantly. In addition, the BaBar Collaboration observed two additional states~\cite{BaBar:2007zjt}, $\Xi_c(3055)$ and $\Xi_c(3123)$, decaying to the $\Lambda_cK\pi$ final state through intermediate $\Sigma_c(2455)$ or $\Sigma_c(2520)$ state. Subsequently, the Belle collaboration confirmed the former state $\Xi_c(3055)$~\cite{Belle:2013htj}, and observed a new decay channel $\Lambda D$ for it~\cite{Belle:2016tai}. As for the latter state, $\Xi_c(3123)$, it has not yet been confirmed by other experimental groups.
Recently, the LHCb collaboration further determined the spin-parity of the $\Xi_c(3055)$ to be $J^P=3/2^+$, and suggested that it is most likely the first $D$-wave $\lambda$-mode $\Xi_c$ excitation~\cite{LHCb:2024eyx}. This also provides strong constraints on the internal properties of other excited states. Furthermore, a new state $\Xi_c(2882)$ was observed by the LHCb collaboration in 2023~\cite{LHCb:2022vns}. This state has not yet been confirmed by other experimental groups, and hence the data are very limited. To date, there are twenty-two $\Xi_c$ and $\Xi_c'$ baryons observed experimentally~\cite{ParticleDataGroup:2026mpi} and we collected the excited states in Table~\ref{Data}.

\begin{table*}[]
	\caption{\label{Data} The mass, width, and observed strong decay mode of the $\Xi_c$ and $\Xi_c'$ baryons in experiments. The unit is MeV.}
	\centering
	\begin{tabular}{cclccclc}
		\hline\hline
State     &$J^P$  &Mass \& Width &Observed strong decay mode  &State     &$J^P$  &Mass \& Width &Observed strong decay mode \\  \hline
$\Xi_c(2790)^+$ &$1/2^-$ &M=$2792.0\pm0.5$  &$\Xi_c'\pi$  &$\Xi_c(2965)^0$ &$?^?$ &M=$2964.88\pm0.33$  &$\Lambda_cK$\\
          &        &$\Gamma=8.9\pm0.6\pm0.8$                  & & &        &$\Gamma=14.1\pm0.9\pm1.3$   &  \\
$\Xi_c(2790)^0$ &$1/2^-$ &M=$2794.0\pm0.5$  &$\Xi_c'\pi$, $\Lambda_c K$ &$\Xi_c(2970)^+$ &$1/2^+$ &M=$2965.2\pm1.0$  &$\Lambda_c\bar{K}\pi$, $\Sigma_c\bar{K}$, $\Xi_c\pi\pi$, $\Xi_c'^{(*)}\pi$\\
          &        &$\Gamma=10.0\pm0.7\pm0.8$                  &  &  &      &$\Gamma=27.2^{+2.8}_{-3.3}$   & \\		
$\Xi_c(2815)^+$ &$3/2^-$ &M=$2816.60\pm0.16$  &$\Xi_c'^{(*)}\pi$, $\Xi_c\pi\pi$ &$\Xi_c(2970)^0$ &$1/2^+$ &M=$2967.3\pm1.9$  &$\Lambda_c\bar{K}\pi$, $\Sigma_c\bar{K}$, $\Xi_c\pi\pi$, $\Xi_c'^{(*)}\pi$\\
                 &        &$\Gamma=2.15\pm0.15$   & & &        &$\Gamma=30.3\pm2.3^{+1.0}_{-1.8}$   &  \\
$\Xi_c(2815)^0$ &$3/2^-$ &M=$2819.85\pm0.27$  &$\Xi_c'^{(*)}\pi$ &$\Xi_c(3055)^+$ &$3/2^+$ &M=$3055.2\pm0.7$  &$\Sigma_cK$, $\Lambda D$\\
                 &        &$\Gamma=2.54\pm0.18\pm0.17$   & & &        &$\Gamma=8.8\pm0.8$   &\\
$\Xi_c(2882)^0$ &$?^?$ &M=$2881.8\pm9.0$  &$\Lambda_cK$ &$\Xi_c(3055)^0$ &$3/2^+$ &M=$3061.00\pm0.83$  &$\Sigma_cK$, $\Lambda D$\\
                 &        &$\Gamma=12.4\pm5.3\pm5.8$   && &        &$\Gamma=8.8\pm0.8$  & \\
$\Xi_c(2923)^+$ &$?^?$ &M=$2922.8\pm0.6$  &$\Xi_c\pi\pi$ &$\Xi_c(3080)^+$ &$?^?$ &M=$3077.2\pm0.4$  &$\Lambda_c\bar{K}\pi$, $\Sigma_c^{(*)}\bar{K}/K$, $\Xi_c\pi\pi$, $\Lambda D$\\
                 &        &$\Gamma=5.3\pm0.9\pm1.4$   & &&        &$\Gamma=3.8\pm0.9$   & \\
$\Xi_c(2923)^0$ &$?^?$ &M=$2923.2\pm0.4$  &$\Lambda_cK$ &$\Xi_c(3080)^0$ &$?^?$ &M=$3079.9\pm1.4$  &$\Lambda_c\bar{K}\pi$, $\Sigma_c^{(*)}\bar{K}$\\
                 &        &$\Gamma=5.8\pm1.3$   & & &        &$\Gamma=5.6\pm2.2$   & \\
$\Xi_c(2939)^+$ &$?^?$ &M=$2942.3\pm4.6$  &$\Lambda_cK$ &$\Xi_c(3123)^+$ &$?^?$ &M=$3122.9\pm1.3$  &$\Sigma_cK$\\
                 &        &$\Gamma=14.8\pm8.8\pm2.5$   & &  &        &$\Gamma=4.4\pm3.4\pm1.7$   &\\
$\Xi_c(2939)^0$ &$?^?$ &M=$2938.55\pm0.30$  &$\Lambda_cK$\\
                &        &$\Gamma=10.2\pm0.8\pm1.1$   & \\
\hline\hline
\end{tabular}
\end{table*}

The spectroscopy and strong decays of excited $\Xi_c$ and $\Xi'_c$ baryons have attracted considerable theoretical interest in recent years. A variety of theoretical approaches, including constituent quark models, heavy-quark effective theory (HQET), QCD sum rules (QCDSR), lattice QCD, and effective field theories, have been employed to investigate their mass spectra and internal structures, while the ${}^3P_0$ pair-creation model, chiral quark models, QCDSR/light-cone QCD sum rules (LCSR), and coupled-channel approaches have been widely applied to study their strong-decay properties~\cite{Crede2024,Luo:2025sns,Capstick1986,Roberts1992,Yang:2020zjl,Agaev:2020fut,Lu:2020ivo,Liu2012,Zhao2016,Chen2016}. In the conventional three-quark picture, excited $\Xi_c$ baryons can be classified according to their flavor multiplets, radial and orbital excitations, and $\rho$- and $\lambda$-mode configurations. Among the established low-lying states, the $\Xi_c(2790)$ and $\Xi_c(2815)$ are generally interpreted as the $1P_\lambda$ antitriplet states with $J^P=1/2^-$ and $3/2^-$, respectively~\cite{Crede2024,Capstick1986,Roberts1992,Yang:2020zjl,Agaev:2020fut,Lu:2020ivo}. However, the nature of several higher excited states remains unsettled. In particular, the $\Xi_c(2882)$, $\Xi_c(2923)$, $\Xi_c(2939)$, and $\Xi_c(2965)$ have been assigned to different $1P$, $\rho/\lambda$-mode, and $2S$ configurations with various quantum numbers in different quark-model, decay-model, and QCDSR analyses~\cite{Luo:2025sns,Lu:2020ivo,Yang:2020zjl,Agaev:2020fut,Hu:2025brq,Zhang:2026ugd,LuoXic2026}. The $\Xi_c(2970)$, whose spin has been experimentally determined to be $J=1/2$~\cite{Belle:2020tom}, is commonly interpreted as a $2S$ excitation with $J^P=1/2^+$, although alternative configurations have also been discussed~\cite{Zhao:2020tpf,Chen:2016phw,Chen:2015kpa,Chen:2007xf,Bahtiyar:2020uuj}. The $\Xi_c(3055)$ has recently received particular attention following the first experimental determination of its spin and parity as $J^P=3/2^+$, which supports its interpretation as the first $D$-wave $\lambda$-mode excitation of the $\Xi_c$ flavor triplet~\cite{LHCb:2024eyx}. The $\Xi_c(3080)$ is also commonly considered to be a $D$-wave excitation, while the nature of the higher-mass $\Xi_c(3123)$ remains poorly established and several $D$-wave or higher-excitation configurations have been proposed~\cite{Liu2012,Zhao2016,Chen2016}.

Beyond the conventional three-quark picture, non-conventional interpretations, including hadronic molecular configurations and coupled-channel dynamics, have also been proposed for some excited $\Xi_c$ and $\Xi'_c$ states~\cite{Yu:2018yxl,Zhu:2020jke}. These competing interpretations, together with the different quark-model assignments, indicate that the mass spectrum alone may not be sufficient to determine the internal structures of the excited $\Xi_c$ and $\Xi'_c$ baryons. Strong-decay properties, such as total and partial decay widths and decay-width ratios, provide complementary information that is sensitive to the underlying wave functions and quantum numbers. Therefore, a combined analysis of the mass spectrum and strong-decay properties provides a more effective approach to discriminate among different assignments and to clarify the nature of the observed excited $\Xi_c$ and $\Xi'_c$ states. In this work, we first investigate the mass spectrum of the $1S$-, $1P$- and $1D$-wave $\Xi_c$ and $\Xi'_c$ states with potential quark model to identify the most plausible spectroscopic configurations, and then analyze their strong-decay properties within the quark pair creation model to further constrain the assignments and explore the internal structures of the observed controversial exited states.

This paper is structured as follows. In Sec. II, we
introduce the potential quark model, quark pair creation model and the relationship of states in different coupling schemes.
Then we present the masses and strong decay properties of the $\Xi_c$ and $\Xi'_c$ baryons, while discuss the possible assignment of the observed $\Xi_c$ and $\Xi'_c$ baryons in Sec. III. A brief summary is given in Sec. IV.

\section{Theoretical framework}

\subsection{Mass spectrum}

\subsubsection{Hamiltonian}

  Numerous models and methods have been developed to compute hadron mass spectra, including lattice QCD, QCD sum rules, relativistic bag models, Regge theory, and nonrelativistic quark potential models.The quark potential model serves as one of the effective approaches for investigating nonperturbative effects in low-energy QCD. Godfrey, Isgur et al. later extended the nonrelativistic quark potential model into a relativistic framework~\cite{Mass.theo.1,Mass.theo.2}. The core procedure involves solving the Schr\"odinger equation to extract masses and realistic wave functions. The relativistic quark potential model possesses two distinctive merits: (1) it enjoys a broader applicability and can treat mesons and baryons in a unified manner~\cite{Mass.theo.18,Mass.theo.19,Mass.theo.20,Mass.theo.21}; (2) it has weak parameter dependence.

 In this work, we focus on the relativistic quark potential model proposed by Godfrey and Isgur, and compute the mass spectra of $\Xi_c$ and $\Xi_c'$ within this model.
The Hamiltonian of the Schr\"odinger equation is composed of two components: the relativistic kinetic term $T$ and various potential terms $V$, which reads
 \begin{equation}
 	H_{0}=\sum_{i=1}^{3}\sqrt{\vec{\mathbf{p}}_{i}^{\,2}+m_{i}^{2}}+\sum_{i<j}^{3}\left(V_{ij}+C_{ij}\right)
 \end{equation}
 where $\sqrt{\vec{\mathbf{p}}_{i}^{\,2}+m_{i}^{2}}$ corresponds to the relativistic kinetic term $T_i$ of the $i$-th quark.
 $V_{ij}$ and $C_{ij}$ denote the effective potential and zero-point energy between the $i$-th and $j$-th quarks, respectively. The zero-point energy is a constant fitted from well-established experimental and theoretical states, with its values presented in Table~\ref{Table:Parameters} below.

 The effective potential consists of contributions from long-range and non-long-range interactions. The long-range component is referred to as the confining potential $V_{\mathrm{conf}}$. The non-long-range part mainly contains two types of potentials: the short-range one-gluon exchange potential (OGE)~\cite{Mass.theo.1,Mass.theo.2,Mass.theo.3,Mass.theo.4,Mass.theo.5}, and the intermediate-short-range one-boson exchange potential (OBE)~\cite{Mass.theo.6,Mass.theo.7,Mass.theo.8}. In the present calculation, we employ the conventional linear confining potential together with the Isgur--Karl one-gluon exchange potential $V_{\mathrm{OGE}}$ for mass-spectrum evaluation. Consequently, the total potential can be written as

\begin{equation}
	V_{ij}=V_{\mathrm{conf}}+V_{\mathrm{OGE}}.
\end{equation}
The confinement term $V_{\rm conf}$, which describes the long-range interaction, accounts for the color confinement effect between quarks and is expressed as
\begin{equation} V_{\mathrm{conf}}=\sum_{i<j}-\frac{3}{16}\left(\vec{\lambda}_{i}\cdot\vec{\lambda}_{j}\right)\cdot\left[b_{ij}f(r_{ij})\right].
\end{equation}
Here $\vec{\lambda}_{i}\cdot\vec{\lambda}_{j}$ denotes the color operator, and $b_{ij}$ is the parameter for the strength of the confining potential between the $i$-th and $j$-th quarks. $f(r_{ij})$ stands for the two-body confining potential, and exists various forms as follows~\cite{Mass.theo.3,Mass.theo.5,Mass.theo.9,Mass.theo.10}:
\begin{equation}
	f(r_{ij})=
	\begin{cases}
		r_{ij} & \text{linear potential}\\[4pt]
		(1-e^{-\mu r_{ij}})/r_{ij} & \text{screened potential}\\[4pt]
		r_{ij}^{2} & \text{harmonic oscillator potential}\\[4pt]
		\mathrm{erf}(r_{ij}) & \text{error-function potential}
	\end{cases}
\end{equation}
In this work, we adopt the linear confining potential to describe the color confinement between quarks.

For the one-gluon exchange potential $V_{\mathrm{OGE}}$, it consists of the Coulomb term $V_{ij}^{\mathrm{Coul}}$ and spin-dependent terms. The spin-dependent terms are further divided into three parts: the spin-spin term $V_{ij}^{ss}$, the tensor potential $V_{ij}^{T}$, and the spin-orbit term $V_{ij}^{Ls}$. Hence,
\begin{equation}
	V_{\mathrm{OGE}}=\sum_{i<j}V_{ij}^{\mathrm{Coul}}+V_{ij}^{SS}+V_{ij}^{LS}+V_{ij}^{T}.
\end{equation}
The explicit forms of each potential are presented below.

\textbf{a.} \quad Coulomb potential $V_{ij}^{\mathrm{Coul}}$:
\begin{equation}
	V_{ij}^{\mathrm{Coul}}=\sum_{i<j}-\frac{2\alpha_{ij}}{3}\frac{1}{r_{ij}},
\end{equation}
where $\alpha_{ij}$ is the coupling constant for one-gluon exchange.

\textbf{b.} \quad Spin-spin term $V_{ij}^{ss}$:
\begin{equation}
	V_{ij}^{ss}=\sum_{i<j}-\frac{\alpha_{ij}}{4}\left(\vec{\lambda}_{i}\cdot\vec{\lambda}_{j}\right)
	\left\{
	\frac{\pi}{2}\cdot\frac{\sigma_{ij}^{3}e^{-\sigma_{ij}^{2}r_{ij}^{2}}}{\pi^{\frac{3}{2}}}\cdot\frac{16}{3m_{i}m_{j}}\left(\vec{S}_{i}\cdot\vec{S}_{j}\right)
	\right\},
\end{equation}
where $\vec{S}_{i}\cdot\vec{S}_{j}$ represents the spin operator, and $\sigma_{ij}$ is a spin-related constant.

\textbf{c.} \quad Tensor potential $V_{ij}^T$:
\begin{equation}
	\begin{aligned}
		V_{ij}^T &= \sum_{i<j}-\frac{\alpha_{ij}}{4}\left(\vec{\lambda}_i\cdot\vec{\lambda}_j\right)
		\frac{1}{m_i m_j r_{ij}^3}
		\left\{
		\frac{3\left(\vec{S}_i\cdot\vec{r}_{ij}\right)\left(\vec{S}_j\cdot\vec{r}_{ij}\right)}{r_{ij}^2}
		-\vec{S}_i\cdot\vec{S}_j
		\right\}.
	\end{aligned}
	\label{eq:tensor_pot}
\end{equation}
Since $\dfrac{\vec{r}}{r}=\hat{\boldsymbol{r}}$, the tensor term operator $\dfrac{\left(\vec{S}_{i}\cdot\vec{r}_{ij}\right)\left(\vec{S}_{j}\cdot\vec{r}_{ij}\right)}{r_{ij}^{2}}$ can be rewritten as $\left(\vec{S}_{i}\cdot\hat{\boldsymbol{r}}\right)\left(\vec{S}_{j}\cdot\hat{\boldsymbol{r}}\right)$. Then, the operator $\vec{S}\cdot\hat{\boldsymbol{r}}$ expands as
\begin{equation}
	\boldsymbol{S}\cdot\hat{\boldsymbol{r}}=\frac{1}{2}\hat{r}_{+}S_{-}+\frac{1}{2}\hat{r}_{-}S_{+}+\hat{r}_{z}S_{z}.
\end{equation}

\textbf{d.} \quad Spin-orbit term $V_{ij}^{Ls}$:

The spin-orbit term $V_{ij}^{Ls}$ comprises the color-magnetic term
\begin{equation}
	\begin{aligned}
		V_{ij}^{\mathrm{so(v)}}
		&= \sum_{i<j} \frac{1}{r_{ij}} \frac{dV_{ij}^{\mathrm{Coul}}}{dr_{ij}}
		\left(
		\frac{\vec{r}_{ij}\times\vec{p}_i\cdot\vec{S}_i}{2m_i^2}
		-\frac{\vec{r}_{ij}\times\vec{p}_j\cdot\vec{S}_j}{2m_j^2}
		\right. \\
		&\quad \left.
		-\frac{\vec{r}_{ij}\times\vec{p}_j\cdot\vec{S}_i - \vec{r}_{ij}\times\vec{p}_i\cdot\vec{S}_j}{m_i m_j}
		\right),
	\end{aligned}
	\label{eq:so_v}
\end{equation}
and the Thomas-precession term
\begin{equation}
	V_{ij}^{so(TP)}=\sum_{i<j}-\frac{1}{r_{ij}}\frac{\partial V_{ij}^{\mathrm{conf}}}{\partial r_{ij}}
	\left(
	\frac{\vec{r}_{ij}\times\vec{p}_{i}\cdot\vec{S}_{i}}{2m_{i}^{2}}
	-\frac{\vec{r}_{ij}\times\vec{p}_{j}\cdot\vec{S}_{j}}{2m_{j}^{2}}
	\right).
\end{equation}

\subsubsection{Wave Function}

The wave function used to solve the Schr\"odinger equation consists of four components: the spatial wave function $\psi$, the spin wave function $\chi$, the color wave function $\zeta^c$, and the flavor wave function $\phi$. The total wave function is the direct product of these components,
\[
\Psi=\psi \otimes \chi \otimes \zeta^c \otimes \phi.
\]
In the mass spectrum calculation, since there exists no operator related to the flavor wave function $\phi$, the flavor wave function is directly normalized out.

The evaluation of spatial matrix elements is the core of solving the mass spectrum. Here, the radial spatial wave function adopts the single-particle explicitly correlated Gaussian basis as the basis set~\cite{Mass.theo.11,Mass.theo.12,Mass.theo.13}, whose form reads
\begin{equation}
	\psi_R(r_1,r_2,r_3)=\exp\left[-\sum_{i<j}^{3}\alpha_{ij}r_{ij}^{2}\right]
	=\exp\left[-f r_{12}^{2}-d r_{13}^{2}-e r_{23}^{2}\right].
\end{equation}
where $\alpha_{ij}$ denotes the variational parameter between the $i$-th and $j$-th quarks. The $\Xi_c$ baryon contains one u/d quark, one s quark and one c quark. Therefore, three Gaussian basis parameters are introduced in the Gaussian basis to describe different quark pairs: $\alpha_{12}=f$ for quark pair $12$, $\alpha_{13}=d$ for quark pair $13$, and $\alpha_{23}=e$ for quark pair $23$.

For the determination of variational parameters $f$, $d$, and $e$, we follow the prescription of Hiyama et al., where the variational parameters are taken as geometric sequences~\cite{Mass.theo.22,Mass.theo.15},
\begin{equation}
	f=\frac{1}{\left(df_1\times af^{n-1}\right)^2},\quad
	d=\frac{1}{\left(dd_1\times ad^{m-1}\right)^2},\quad
	e=\frac{1}{\left(de_1\times ae^{v-1}\right)^2}.
\end{equation}
Here $n$, $m$, and $v$ are the numbers of terms selected for the three geometric sequences. The total number of Gaussian wave functions is determined by $n\times m\times v$, and $af$, $ad$, $ae$ are the common ratios of the corresponding geometric sequences. Therefore, the parameters to be determined in our calculation include the starting variational parameters $df_1$, $dd_1$, $de_1$, the ending variational parameters $df_n$, $dd_m$, $de_v$, and the number of Gaussian basis functions.

To simplify the calculation, the coordinates of the three particles $r_1$, $r_2$, $r_3$ are transformed into the Jacobi coordinates $\rho_{12}$ and $\lambda_{12}$. The transformation relation between them is given below:

\begin{equation}
	\vec{\boldsymbol{\rho}}_{12}=\frac{1}{\sqrt{2}}\left(\vec{\boldsymbol{r}}_1-\vec{\boldsymbol{r}}_2\right)
\end{equation}
\begin{equation}
	\vec{\boldsymbol{\lambda}}_{12}=\sqrt{\frac{2}{3}}\left(\frac{m_1\vec{\boldsymbol{r}}_1+m_2\vec{\boldsymbol{r}}_2}{m_1+m_2}-\vec{\boldsymbol{r}}_3\right)
\end{equation}

\begin{figure}[]
  \centering
  \includegraphics[width=0.3\textwidth]{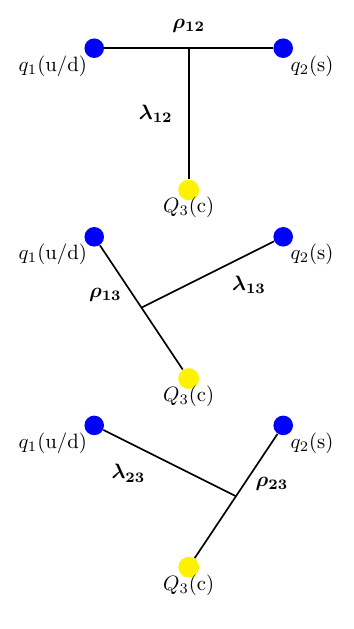}
  \caption{Three sets of Jacobi coordinates for three-quark system of $\Xi_c$/$\Xi_c'$ baryon.}
  \label{Fig:Jacobi}
\end{figure}

It can be seen that $\rho_{12}$ is the relative coordinate between quark 1 and quark 2. In a similar manner, one can also define the Jacobi coordinates $\rho_{13}$ and $\rho_{23}$ for quark pairs $13$ and $23$, together with $\lambda_{13}$ and $\lambda_{23}$~\cite{Mass.theo.11,Mass.theo.14,Mass.theo.23}. Their definitions are shown in the figure~\ref{Fig:Jacobi}.
	
In this way, the three-body problem for the spatial wave function is reduced to a relatively simple two-body problem, which greatly simplifies the investigation. After reorganizing the expressions, taking the Jacobi coordinates $\rho_{12}$ and $\lambda_{12}$ associated with quark pair $12$ as an example, the radial spatial wave function reads,

\begin{equation}
	\begin{aligned}
		\psi_R(\rho_{12},\lambda_{12})
		&= 		\exp\big[-aa\,\rho_{12}^2-2bb\,\rho_{12}\lambda_{12}-cc\,\lambda_{12}^2\big]
	\end{aligned}
	\label{eq:wave_gauss}
\end{equation}

Thus, the complete expression for the spatial wave function is given below,

\begin{equation}
	\begin{aligned}
		\psi(\rho_{12},\lambda_{12})
		&= N_{\text{nor}} \rho_{12}^{l_{\rho_{12}}}\lambda_{12}^{l_{\lambda_{12}}}
		\exp\big[-aa\,\rho_{12}^2-2bb\,\rho_{12}\lambda_{12}-cc\,\lambda_{12}^2\big] \\
		&\quad \times Y_{l_{\lambda_{12}}m_{\lambda_{12}}} Y_{l_{\rho_{12}}m_{\rho_{12}}}.
	\end{aligned}
	\label{eq:wave_gauss}
\end{equation}
Here, $\rho^{l_\rho}\lambda^{l_\lambda}Y_{l_\lambda m_\lambda}Y_{l_\rho m_\rho}=y_{l_\lambda m_\lambda}y_{l_\rho m_\rho}$ corresponds to the angular part. The prefactors $aa$, $bb$, $cc$ appearing before the Jacobi coordinates are expressions related to the Gaussian basis parameters $f$, $d$, $e$ and the masses of the three quarks $m_1$, $m_2$, $m_3$. Therefore, the coordinate operators in the Hamiltonian also need to be transformed into Jacobi-coordinate form. For the terms involving $r_{13}$ and $r_{23}$ in the Hamiltonian, one may redefine two sets of Jacobi coordinates according to the relation between $r_{12}$ and Jacobi coordinates. The procedure is identical to the way we define different Jacobi coordinates for wave functions in the previous context.

To facilitate the evaluation of the relativistic kinetic energy term, the form of the relativistic kinetic energy term needs to be simplified. In the three-quark baryon system with three distinct quark flavors,
\begin{equation}
	\vec{P}_{\lambda 12}=\sqrt{\frac{3}{2}}\cdot \frac{m_{3}\cdot (\vec{P}_{1}+\vec{P}_{2})-(m_{1}+m_{2})\cdot \vec{P}_{3}}{m_{1}+m_{2}+m_{3}},
\end{equation}
\begin{equation}
	\vec{P}_{\lambda 13}=\sqrt{\frac{3}{2}}\cdot \frac{m_{2}\cdot (\vec{P}_{1}+\vec{P}_{3})-(m_{1}+m_{3})\cdot \vec{P}_{2}}{m_{1}+m_{2}+m_{3}},
\end{equation}
\begin{equation}
	\vec{P}_{\lambda 23}=\sqrt{\frac{3}{2}}\cdot \frac{m_{1}\cdot (\vec{P}_{2}+\vec{P}_{3})-(m_{2}+m_{3})\cdot \vec{P}_{1}}{m_{1}+m_{2}+m_{3}}.
\end{equation}
Together with momentum conservation
\begin{equation}
	\vec{P}=\vec{P}_{1}+\vec{P}_{2}+\vec{P}_{3}=0,
\end{equation}
the relativistic kinetic term $T=\sum_{i=1}^{3}\sqrt{\vec{\mathbf{p}}_{i}^{\,2}+m_{i}^{2}}$ can be cast into the form
\begin{equation}
	T=\sqrt{\frac{2}{3}\vec{P}_{\lambda 23}^{\,2}+m_{1}^{2}}+\sqrt{\frac{2}{3}\vec{P}_{\lambda 13}^{\,2}+m_{2}^{2}}+\sqrt{\frac{2}{3}\vec{P}_{\lambda 12}^{\,2}+m_{3}^{2}}.
\end{equation}
Here $\mathrm{P}_{\lambda 12}$, $\mathrm{P}_{\lambda 13}$ and $\mathrm{P}_{\lambda 23}$ denote the momenta associated with Jacobi coordinates. Meanwhile, the spatial wave function also needs to be Fourier-transformed from coordinate space to momentum space.

To solve the Schr\"odinger equation, it should be noted that the overlap matrix $N$ must be introduced in order to evaluate the generalized matrix elements(see Eq.~\ref{23} for its definition). One solves the eigenvalue or generalized eigenvalue problem for $H$ and $N$, and the minimal mass obtained from the variational method corresponds to the mass of a given state. The wave function is expanded in multiple Gaussian basis functions, where each basis component $\psi_i$ is multiplied by an expansion coefficient $C_i$. In the present work, the number of variational basis terms is set to $5$, so the total physical wave function is a superposition of $5\times5\times5=125$ Gaussian basis functions. When $n$, $m$, and $v$ are all equal to $1$, the corresponding expansion coefficient is $C_1$. The final physical wave function can be written as $\Psi=C_1\psi_1+\dots+C_{125}\psi_{125}$.

The procedure for solving the Schr\"odinger equation is as follows. We first construct each term of the Hamiltonian, then left-multiply by $\psi$. Therefore, the core task is to compute the matrix elements of kinetic and potential terms as well as the overlap matrix elements. These matrices are of dimension $125\times125$,
\begin{equation}\label{23}
	\langle \Psi_f \vert \hat{H} \vert \Psi_i \rangle = H_{fi},\quad
	\langle \Psi_f \vert \Psi_i \rangle = N_{fi}.
\end{equation}
Hence, solving the Schr\"odinger equation for the mass spectrum reduces to solving the generalized eigenvalue equation:
\begin{equation}
	\sum_{i=1}^{N_{\text{max}}}\left(H_{fi}-E N_{fi}\right)C_{i}=0.
\end{equation}
The lowest eigenvalue and its corresponding eigenfunction yield the mass and physical wave function of the relevant state.

\subsubsection{parameters}

The quark masses are fixed as $m_u=m_d=0.45\ \mathrm{GeV}$, $m_s=0.55\ \mathrm{GeV}$, and $m_c=1.45\ \mathrm{GeV}$. The other parameters, including the confinement potential parameter $b_{ij}$, vacuum zero-point energy $C_{ij}$, coupling constant $\alpha_{ij}$, and spin-dependent constant $\sigma_{ij}$, are fitted to experimentally well-established states: the ground states $\Xi_c$, $\Xi_c'$, and $\Xi_c'(2645)$, and the excited states $\Xi_c(2790)$ and $\Xi_c(2815)$. The fitted values of $b_{ij}$ are consistent with those reported in the literature, lying in the range $0.10$--$0.14\ \mathrm{GeV}^2$~\cite{Mass.theo.16,Mass.theo.17}. The values of $\sigma_{ij}$ are taken from Ref.~\cite{Mass.theo.15}.

It should be noted that the spin-orbit and tensor terms contain the factor $1/r_{ij}^3$, which may lead to unstable solutions during numerical calculations. To obtain stable solutions, we follow the approach in the literature by introducing a cutoff parameter $r_c$ for these two terms and adopt $r_{c(\mathrm{sc})}=0.16\ \mathrm{fm}$ as in Refs.~\cite{Mass.theo.19,Mass.theo.20}.

\begin{table}[htbp]
	\centering
	\small
	\renewcommand{\arraystretch}{1.4}
	\setlength{\tabcolsep}{5pt}
	\begin{tabular}{*{4}{c}}
		\noalign{\hrule height 1pt}
		$m_{u/d}(\mathrm{GeV})$ & $m_s(\mathrm{GeV})$ & $m_c(\mathrm{GeV})$ & $b(\mathrm{GeV}^2)$ \\
		0.45 & 0.55 & 1.45 & 0.12 \\
		\hline
		$C_{0(u/ds)}(\mathrm{GeV})$ & $C_{0(u/dc)}(\mathrm{GeV})$ & $C_{0(sc)}(\mathrm{GeV})$ & $\alpha_{u/ds}$ \\
		$-0.072$ & $-0.125$ & $-0.084$ & 0.86 \\
		\hline
		$\alpha_{u/dc}$ & $\alpha_{sc}$ & $\sigma_{u/ds}(\mathrm{GeV})$ & $\sigma_{u/dc}(\mathrm{GeV})$ \\
		0.6 & 0.56 & 0.52 & 0.62 \\
		\hline
		$\sigma_{sc}(\mathrm{GeV})$ & $r_{c(u/ds)}(\mathrm{fm})$ & $r_{c(u/dc)}(\mathrm{fm})$ & $r_{c(sc)}(\mathrm{fm})$ \\
		0.68 & 0.2 & 0.2 & 0.16 \\
		\noalign{\hrule height 1pt}
	\end{tabular}
	\caption{Parameters of mass spectrum}\label{Table:Parameters}
\end{table}

   As mentioned above, we set the variational parameters $n=m=v=5$. In addition, the starting values for variation are $df_1=0.1\ \mathrm{fm}$, $dd_1=0.2\ \mathrm{fm}$, $de_1=0.2\ \mathrm{fm}$, and the ending values for variation are $df_n=2\ \mathrm{fm}$, $dd_n=3\ \mathrm{fm}$, $de_n=3\ \mathrm{fm}$. Such parameter choices yield stable solutions, and most of the subsequent calculated root-mean-square radii are reasonably well-behaved.

\subsection{The quark pair creation model}

\begin{figure}[]
  \centering
  \includegraphics[width=0.45\textwidth]{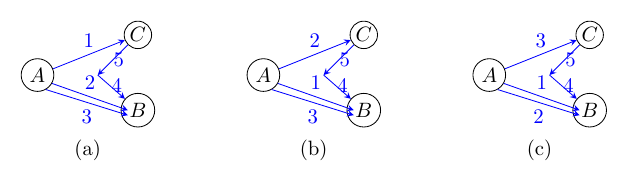}
  \caption{Schematic diagram for the decay of initial baryon $A$ into final-state baryon $B$ and meson $C$.}
  \label{Fig:QPCModel}
\end{figure}

  The quark pair creation (QPC) model plays an important role in investigating the strong decay behaviors of mesons and baryons. It was first proposed by Micu, Carlitz and Kislinger~\cite{3P0.Meth.1,3P0.Meth.2}, and further developed by the Le Yaouanc group~\cite{3P0.Meth.3,3P0.Meth.4,3P0.Meth.5}. Within this model, the strong decay takes place via the creation of quark-antiquark pair from the vacuum with quantum number $0^{++}$. This quark-antiquark pair then undergoes quark rearrangement with the quarks inside the initial-state particle, decaying into two final-state particles. For the $\Xi_c^+$ baryons, which are composed of two light quarks($s$ and $u$) and a heavy quark $c$, the quark rearrangement process proceeds as follows: one quark from the initial baryon state $|A\rangle$ regroups with the created antiquark to form a meson $|C\rangle$, while the remaining two quarks regoroup with the created quark to form a daughter baryon $|B\rangle$, as shown in Figure~\ref{Fig:QPCModel}.

The transition operator for this decay process in the nonrelativistic limit is given by,

\begin{align}
	T &= -3\gamma \sum_{m} \langle 1m;1-m|00\rangle
	\int d^3\mathbf{p}_4 d^3\mathbf{p}_5 \delta^3(\mathbf{p}_4+\mathbf{p}_5) \nonumber\\
	&\quad \times Y_1^{m}\left(\frac{\mathbf{p}_4-\mathbf{p}_5}{2}\right)
	\chi^{45}_{1-m}\phi^{45}_0 \omega^{45}_0
	a^\dagger_{4i}(\mathbf{p}_4) b^\dagger_{5j}(\mathbf{p}_5).
\end{align}
The dimensionless parameter $\gamma$ characterizes the strength of vacuum pair creation.
$\mathbf{p}_4$ and $\mathbf{p}_5$ are the three-dimensional momenta of the quark pair created from the vacuum. $\mathcal{Y}_1^m=|\mathbf{p}|Y_1^m(\theta_p,\phi_p)$ denotes the solid harmonic polynomial, which describes the momentum-space distribution.
$\chi_{1-m}^{45}$, $\phi_0^{45}=(\mathrm{u}\bar{\mathrm{u}}+\mathrm{d}\bar{\mathrm{d}}+\mathrm{s}\bar{\mathrm{s}})/\sqrt{3}$, and $\omega_0^{45}=\delta_{ij}$ correspond to the spin triplet, flavor wave function, and color singlet, respectively. $a_{4i}^\dagger b_{5j}^\dagger$ is the creation operator representing quark pair production from the vacuum.

The spatial wave function of the initial-state baryon is taken as the real wave function from the mass-spectrum calculation; for simplicity, however, it is replaced by an effective harmonic-oscillator wave function. The spatial wave functions of the final-state baryon and meson are described by simple harmonic-oscillator wave functions.
Then, we can obtain the transition amplitude in the center of mass frame,
\begin{equation}\label{eq:M_transition}
	\begin{gathered}
		\mathcal{M}^{M_{J_A}M_{J_B}M_{J_C}}(A\to BC) ~~~~~~~~~~~~~~~~~~~~~~~~~~~~~~~~~~~~~~~~~~~~~~~~~~~~~~~~~~~\\
		=\gamma\sqrt{8E_A E_B E_C}\displaystyle\prod_{A,B,C}
		\left\langle \chi^{124}_{S_B M_{S_B}} \chi^{35}_{S_C M_{S_C}}
		\bigg|\chi^{123}_{S_A M_{S_A}} \chi^{45}_{1-m}\right\rangle \\
		\times
		\left\langle \varphi^{124}_B \varphi^{35}_C
		\bigg|\varphi^{123}_A \varphi^{45}_0\right\rangle
		I^{M_{L_A},m}_{M_{L_B},M_{L_C}}(\mathbf{p}).
	\end{gathered}
\end{equation}
Here, $\mathrm{I}^{M_{L_\mathrm{A}},\,m}_{M_{L_\mathrm{B}},\,M_{L_\mathrm{C}}}$ denotes the spatial integral. $\prod_{A,B,C}$ represents the Clebsch–Gordan coefficients and its expansion reads,

\begin{align}
	\prod_{A,B,C}
	&=\sum
	\langle L_\mathrm{B} M_{L_\mathrm{B}}; S_\mathrm{B} M_{S_\mathrm{B}} \big| J_\mathrm{B}, M_{J_\mathrm{B}} \rangle
	\langle L_\mathrm{C} M_{L_\mathrm{C}}; S_\mathrm{C} M_{S_\mathrm{C}} \big| J_\mathrm{C}, M_{J_\mathrm{C}} \rangle \notag \\
	&\quad \times
	\langle L_\mathrm{A} M_{L_\mathrm{A}}; S_\mathrm{A} M_{S_\mathrm{A}} \big| J_\mathrm{A}, M_{J_\mathrm{A}} \rangle
	\langle 1\,m;\,1\,-m \big| 0\,0 \rangle.
	\label{eq:2.xx}
\end{align}

While, the transition operator $T$ is effective in the low-momentum region. To suppress the nonphysical contributions in high-momentum region, we introduce a suppressed form factor $e^{-\frac{\mathbf{p}^2}{2\Lambda^2}}$ into the decay amplitude~\cite{3P0.Meth.6,3P0.Meth.7,3P0.Meth.8,3P0.Meth.9,3P0.Meth.11,3P0.Meth.12,3P0.Meth.13},
\begin{align}
	\mathcal{M}^{M_{J_A}M_{J_B}M_{J_C}}(A\to BC)
	\to
	\mathcal{M}^{M_{J_A}M_{J_B}M_{J_C}}(A\to BC)
	e^{-\frac{\mathbf{p}^2}{2\Lambda^2}},
\end{align}
as down in our previous work~\cite{Zhang:2025rqe}. Here,
$\mathbf{p}$ denotes the momentum of the final-state baryon $B$ in the center-of-mass frame of the initial baryon $A$. In this equation, we adopt the cutoff parameter $\Lambda=780\ \mathrm{MeV}$, consistent with Ref.~\cite{3P0.Meth.10}.

Finally, the relation between the transition amplitude and the width is given by,
\begin{align}
	\Gamma(A\to BC)=\pi^2 \frac{|\mathbf{p}|}{M_A^2}\frac{1}{2J_A+1}
	\sum_{M_{J_A},M_{J_B},M_{J_C}}
	\left|\mathcal{M}^{M_{J_A}M_{J_B}M_{J_C}}e^{-\frac{\mathbf{p}^2}{2\Lambda^2}}\right|^2.
\end{align}

Within the QPC model, the constituent quark masses are consistent with the values of the constituent quark masses used in our mass-spectrum calculation. The vacuum pair-production strength $\gamma=6.65$ is the same as in our previous work~\cite{Zhou:2023xxs}.

\section{Results and Analysis}

In this section, we present the calculated mass spectra and decay properties of $\Xi_c$ and $\Xi_c'$ baryons. First, preliminary analysis is performed based on the theoretical mass spectra. By comparing the predicted masses with experimental measurements, we identify possible internal quantum numbers and excitation modes for the observed resonances. Then, we further carry out cross-validation using the theoretical calculations of decay properties. Combining the information from both mass spectra and decay properties, we finally identify the theoretical candidates corresponding to the experimental resonances. At the same time, we also predict the properties of states that have not yet been observed experimentally, providing theoretical guidance for future experimental searches.

\subsection{Mass Spectra}

Within the potential quark model, the mass spectra of the $1S$-, $1P$- and $1D$-waves single-charm baryons $\Xi_c$ and $\Xi_c'$ are calculated. It should be noted that the $u$ and $d$ quarks differ only in electric charge and are assigned equal masses in the present calculation. Therefore, the isospin mass splittings between $\Xi_c^+/\Xi_c^0$ and $\Xi_c'^+/\Xi_c'^0$ are not considered in our calculations.

\subsubsection{Mass spectrum of the $\Xi_c$ baryons}

\begin{table*}[tp]
	\clearpage
	
	\caption{Mass spectrum of the $\Xi_c$ baryons (in MeV).}
	\label{tab:xic_mass}
	
	\setlength{\tabcolsep}{0.6pt}
	\centering
	
		\begin{tabular}{cccccccccc}
		\noalign{\hrule height 1pt}
		& Mass~\cite{Mass.1} & Mass~\cite{Theo.13} & Mass~\cite{Theo.20} & Mass~\cite{Theo.29} & Mass~\cite{Theo.12} & Mass~\cite{Mass.8} & Ours & PDG & Candidate \\
		$nL|J^P,j\rangle_{\rho/\lambda}$ & RQM(Ebert) & RQM & Three-quark/Quark-diquark & RFT & ChQM & NQM & & & \\
		\hline
		$\Xi_c1\mathrm{S}\big|\tfrac12^+,1\rangle$ & 2476 & 2479 & 2466/2473 & 2470.4 & $2473\pm7$ & 2470 & 2459 & 2467.79 & $\Xi_c^+$ \\
	    $\Xi_c1\mathrm{P}\big|\tfrac12^-,1\rangle_{\lambda}$ & 2792 & 2789 & 2788/2789 & 2785.7 & $2783\pm7$ & 2793 & 2790 & 2792.00 & $\Xi_c(2790)$ \\
		$\Xi_c1\mathrm{P}\big|\tfrac32^-,1\rangle_{\lambda}$ & 2819 & 2820 & 2815/2814 & 2823.9 & $2830\pm7$ & 2820 & 2821 & 2816.60 & $\Xi_c(2815)$ \\	
        $\Xi_c1\mathrm{P}\big|\tfrac12^-,0\rangle_{\rho}$ & & & 2935/- & & $2950\pm4$ & & 2834 &2881.80 & $\Xi_c(2882)?$\\
		$\Xi_c1\mathrm{P}\big|\tfrac12^-,1\rangle_{\rho}$ & & & 2977/- & & $2910\pm3$ & & 2861 &2881.80 & $\Xi_c(2882)?$\\
		$\Xi_c1\mathrm{P}\big|\tfrac32^-,1\rangle_{\rho}$ & & & 2962/- & & $3057\pm5$ & & 3005 & & \\
		$\Xi_c1\mathrm{P}\big|\tfrac32^-,2\rangle_{\rho}$ & & & 3004/- & & $3037\pm5$ & & 3002 & & \\
		$\Xi_c1\mathrm{P}\big|\tfrac52^-,2\rangle_{\rho}$ & & & 3049/- & & $3116\pm8$ & & 3094 & & \\
		$\Xi_c1\mathrm{D}\big|\tfrac32^+,2\rangle_{\lambda\lambda}$ & 3059 & 3061 & 3118/3113 & 3065.9 & $3043\pm8$ & 3033 & 3051 &3055.20 & $\Xi_c(3055)?$ \\
		$\Xi_c1\mathrm{D}\big|\tfrac52^+,2\rangle_{\lambda\lambda}$ & 3076 & 3078 & 3164/3156 & 3093.3 & $3078\pm8$ & 3040 & 3054 &3077.20 & $\Xi_c(3080)?$ \\
		$\Xi_c1\mathrm{D}\big|\tfrac32^+,2\rangle_{\rho\rho}$ & & & 3413/- & & $3353\pm9$ & & 3249 & & \\
		$\Xi_c1\mathrm{D}\big|\tfrac52^+,2\rangle_{\rho\rho}$ & & & 3458/- & & $3373\pm9$ & & 3288 & & \\
		\noalign{\hrule height 1pt}
	\end{tabular}
	\end{table*}

\paragraph*{A. $1S$-wave states}
The central mass of the ground state $\Xi_c^+$ given by the PDG~\cite{ParticleDataGroup:2026mpi} is 2467.95 MeV, and that of $\Xi_c^0$ is 2470.99 MeV. In the present work, the calculated mass of the $\Xi_c$ is 2459.39 MeV(see table~\ref{tab:xic_mass}), which differs from the experimental value by only approximately 10 MeV. Meanwhile, various theoretical predictions for this state lie close to the experimental measurement. Therefore, in terms of the mass spectrum, the ground-state $\Xi_c$ mass is well reproduced.

\paragraph*{B. $1P$-wave states}
The $1P$-wave $\Xi_c$ states contain two $\lambda$-mode excitations and five $\rho$-mode excitations. For single heavy baryon resonances with the same $J^P$ quantum number, the masses of $\rho$-mode excitations are generally larger than those of $\lambda$-mode excitations. This feature originates from the kinematic difference between the two excitation modes: the $\lambda$ mode describes the excitation of the charm quark with respect to the light-quark subsystem, whereas the $\rho$ mode corresponds to the relative motion between the two light quarks. The internal excitation of light quarks contributes a larger kinetic energy, which yields higher masses for $\rho$ configurations under the same quantum numbers. This qualitative rule is consistent with the conclusions of existing constituent quark models.

According to our calculations, the masses of the two lower $1P$-wave $\lambda$-mode $\Xi_c$ states are 2790 MeV and 2821 MeV, respectively. Based on the experimentally observed masses and spin-parity, we naturally interpret $\Xi_c(2790)$ and $\Xi_c(2815)$ as the $1P$-wave $\lambda$-mode $\Xi_c$ excitation with $J^P=1/2^-$ and $J^P=3/2^-$, respectively. This assignment is supported by most mainstream theoretical works. However, some references based on the SU(6)-HQSS model have proposed an alternative viewpoint, arguing that $\Xi_c(2790)$ contains a substantial $\Lambda_c\bar{K}$ hadronic molecular component~\cite{Theo.4}.

For the five higher $1P$-wave $\rho$-mode $\Xi_c$ states, theoretical calculations of their masses remain scarce, and the existing results are still in considerable disagreement(see table~\ref{tab:xic_mass}). The physical origin of this discrepancy remains unclear, so the physical picture of $\rho$-mode excitations still deserves further investigation. By comparing our calculated results with other results, we find that for the two states with spin-parity $1/2^-$, our calculated masses are relatively lower, while for the three states with spin-parity $3/2-$ and $5/2^-$, our results are close to other theoretical results. Considering mass matching, the experimentally observed $\Xi_c(2882)$ is very likely to be the quark configuration $\Xi_c1P|1/2^-,1\rangle_{\rho}$, as our calculated mass for this state is 2861 MeV. Meanwhile, considering the hteoretical uncertainties, the possibility of $\Xi_c(2882)$ being a candidate for $\Xi_c1P|1/2^-,0\rangle_{\rho}$ cannot be ruled out either, since the mass of $\Xi_c1P|1/2^-,0\rangle_{\rho}$ is 2834 MeV. Nevertheless, the final assignment of its quantum number requires further constraints from decay properties, since decay characteristics provide indispensable information for determining the internal structure of resonances.

\paragraph*{C. $1D$-wave states}
The $1D$-wave $\Xi_c$ states include two $\lambda$-mode excitations and two $\rho$-mode excitations. Similarly, the $\rho$-mode excited states are generally heavier than the $\lambda$-mode excited states, by about (200$\sim$300) MeV.

For the two $\lambda$-mode excitations, our predicted masses are about $M\sim3050$ MeV, which are basically consistent with the calculations from most existing theoretical models. The mass of the experimentally observed state $\Xi_c(3055)$ falls exactly in this range. Considering that the spin-parity of $\Xi_c(3055)$ is $3/2^+$, this state is very likely to be $\Xi_c1D|3/2^+,2\rangle_{\lambda\lambda}$. For the other $\lambda$-mode $1D$-wave excited state $\Xi_c1D|5/2^+,2\rangle_{\lambda\lambda}$, considering the theoretical uncertainties, it may ba a candidate for the observed state $\Xi_c(3080)$.

As for the two $\rho$-mode excited states,  the theoretical masses fall roughly in the range of (3300$\sim$3500) MeV,  reflecting considerable uncertainty. Our calculated values are approximately $\sim$3250 MeV, slightly lower than the masses predicted by other models; however, it is basically consistent with them when the model uncertainties are considered.
Since the masses of these states are relatively high, no corresponding resonant signals have been observed experimentally so far. Further investigations are still required to reveal their structures and spectroscopic properties.

\subsubsection{Mass spectrum of the $\Xi_c'$ baryons}
	\begin{table*}[tp]
	\clearpage
	\caption{Mass spectrum of the $\Xi_c'$ baryons (in MeV).}
	\label{tab:xic_prime_mass}
	\setlength{\tabcolsep}{0.6pt}
	\centering
	\begin{tabular}{ccccccccccc}
		\noalign{\hrule height 1pt}
		& Mass~\cite{Mass.1} & Mass~\cite{Theo.13} & Mass~\cite{Theo.18} & Mass~\cite{Theo.20} & Mass~\cite{Theo.29} & Mass~\cite{Theo.12} & Mass~\cite{Mass.8} & Ours ~~~& PDG ~~~~~~~~~~~~& Candidate \\
		$nL|J^P,j\rangle_{\rho/\lambda}$ & RQM(Ebert) & RQM & QCDSR & Three-quark/Quark-diquark & RFT & ChQM & NQM & & & \\
		\hline
		$\Xi_c'1\mathrm{S}\big|\tfrac12^+,1\rangle$ & 2579 & 2589 & & 2571/2577 & 2570.7 & $2565\pm6$ & 2579 & 2564 & 2578.30 & $\Xi_c'$ \\
		$\Xi_c'1\mathrm{S}\big|\tfrac32^+,1\rangle$ & 2649 & 2660 & & 2640/2650 & 2648.2 & $2637\pm6$ & 2649 & 2643 & 2645.17 & $\Xi_c(2645)$ \\
		$\Xi_c'1\mathrm{P}\big|\tfrac12^-,0\rangle_{\lambda}$ & 2936 & 2906 & 2885 & 2893/2893 & 2873.2 & $2874\pm4$ & 2839 & 2882 &2881.80 & $\Xi_c(2882)?$ \\
		$\Xi_c'1\mathrm{P}\big|\tfrac12^-,1\rangle_{\lambda}$ & 2854 & 2912 & 2890 & 2935/2941 & 2886.4 & $2908\pm4$ & 2900 & 2867 &2881.80 & $\Xi_c(2882)?$  \\
		$\Xi_c'1\mathrm{P}\big|\tfrac32^-,1\rangle_{\lambda}$ & 2935 & 2941 & 2929 & 2920/2919 & 2937.9 & $2915\pm5$ & 2921 & 2930 &2922.80 & $\Xi_c(2923)?$ \\
		$\Xi_c'1\mathrm{P}\big|\tfrac32^-,2\rangle_{\lambda}$ & 2912 & 2948 & 2932 & 2962/2966 & 2992.9 & $2924\pm5$ & 2932 & 2948 & 2942.30 & $\Xi_c(2939)?$ \\
		$\Xi_c'1\mathrm{P}\big|\tfrac52^-,2\rangle_{\lambda}$ & 2929 & 2990 & 3000 & 3007/3009 & 3030.5 & $2948\pm7$ & 2927 & 2969 &2965.20 &$\Xi_c(2965)$? \\
        $\Xi_c'1\mathrm{P}\big|\tfrac12^-,1\rangle_{\rho}$ & & 2958 & 2976 & 3040/- & & $3018\pm7$ & & 2946 &2942.30 & $\Xi_c(2939)?$ \\
		$\Xi_c'1\mathrm{P}\big|\tfrac32^-,1\rangle_{\rho}$ & & 3004 & 2989 & 3067/- & & $3038\pm7$ & & 2976 &2965.20 &$\Xi_c(2965)?$  \\
        $\Xi_c'1\mathrm{D}\big|\tfrac12^+,1\rangle_{\lambda\lambda}$ & 3163 & 3177 & & 3238/3239 & 3157.1 & $3136\pm8$ & 3075 & 3153 &3122.90 &$\Xi_c(3123)?$ \\
		$\Xi_c'1\mathrm{D}\big|\tfrac32^+,1\rangle_{\lambda\lambda}$ & 3167 & 3189 & & 3223/3218 & 3189.8 & $3143\pm4$ & 3089 & 3181 & & \\
		$\Xi_c'1\mathrm{D}\big|\tfrac32^+,2\rangle_{\lambda\lambda}$ & 3160 & 3190 & & 3265/3265 & 3207.1 & $3172\pm4$ & 3081 & 3178 & & \\
		$\Xi_c'1\mathrm{D}\big|\tfrac52^+,2\rangle_{\lambda\lambda}$ & 3166 & 3208 & & 3268/3261 & 3236.6 & $3153\pm5$ & 3091 & 3185 & & \\
		$\Xi_c'1\mathrm{D}\big|\tfrac52^+,3\rangle_{\lambda\lambda}$ & 3153 & 3207 & & 3310/3308 & 3279.0 & $3170\pm5$ & 3077 & 3169 & & \\
		$\Xi_c'1\mathrm{D}\big|\tfrac72^+,3\rangle_{\lambda\lambda}$ & 3147 & 3229 & & 3373/3368 & 3304.6 & $3176\pm8$ & 3078 & 3158 & & \\
		$\Xi_c'1\mathrm{D}\big|\tfrac12^+,1\rangle_{\rho\rho}$ & & & & 3532/- & & $3281\pm8$ & & 3055 & & \\
		$\Xi_c'1\mathrm{D}\big|\tfrac32^+,1\rangle_{\rho\rho}$ & & & & 3517/- & & $3336\pm4$ & & 3236 & & \\
		$\Xi_c'1\mathrm{D}\big|\tfrac32^+,2\rangle_{\rho\rho}$ & & & & 3559/- & & $3302\pm4$ & & 3235 & & \\
		$\Xi_c'1\mathrm{D}\big|\tfrac52^+,2\rangle_{\rho\rho}$ & & & & 3563/- & & $3368\pm6$ & & 3298 & & \\
		$\Xi_c'1\mathrm{D}\big|\tfrac52^+,3\rangle_{\rho\rho}$ & & & & 3604/- & & $3362\pm6$ & & 3263 & & \\
		$\Xi_c'1\mathrm{D}\big|\tfrac72^+,3\rangle_{\rho\rho}$ & & & & 3668/- & & $3293\pm9$ & & 3300 & & \\
		\noalign{\hrule height 1pt}
	\end{tabular}
\end{table*}

\paragraph*{A. $1S$-wave states}
The $1S$ configuration of the flavor sextet $\Xi_c'$ baryon contains two sets of spin eigenstates, corresponding to the spin-singlet $\Xi_c'$ and spin-triplet $\Xi_c'^*$ states of the light diquark, respectively. For the spin-singlet ground states $\Xi_c'^+$ and $\Xi_c'^0$, the experimental central masses listed by the PDG are 2578.3 MeV and 2578.8 MeV, respectively. Our theoretical mass for the ground-state is 2564 MeV, with a deviation of approximately 14 MeV from experiment. As for the spin-triplet states $\Xi_c'^{*+}$ and $\Xi_c'^{*0}$, their experimental masses are 2645.17 MeV and 2646.24 MeV, respectively. Since the isospin splitting induced by electric charge is tiny, they are collectively labeled as $\Xi_c(2645)$ by the PDG. Our calculated mass for the spin-triplet state $\Xi_c'^*$ is 2643 MeV, which deviates from the experimental value by only 2 MeV.
According to our calculated results and other theoretical works(see table~\ref{tab:xic_prime_mass}), the $1S$-wave $\Xi_c'$ states can be well explained.

\paragraph*{B. $1P$-wave states}
The $1P$-wave $\Xi_c'$ excited states consist of five $\lambda$-mode excitations and two $\rho$-mode excitations. Based on our calculations and other theoretical studies, the masses of the $\lambda$-mode $\Xi_c'$ resonances fall approximately in the rang of (2900$\sim$3000) MeV, and the masses of the experimentally observed states $\Xi_c(2882)$, $\Xi_c(2923)$, $\Xi_c(2939)$, $\Xi_c(2965)$ and $\Xi_c(2970)$ fall exactly in this range. Therefore, investigating the properties of these $\lambda$-mode excited states is crucial for clarifying the nature of those states. Meanwhile, we notice that the spin-parity of $\Xi_c(2970)$ have been determined to be $1/2^+$ by the Belle Collaboration~\cite{Belle:2020tom}. Hence, the possibility that $\Xi_c(2970)$ corresponds to the $1P$-wave $\lambda$-mode $\Xi_c'$ excitation can be essentially excluded.

According to our calculations, the observed state $\Xi_c(2882)$ has a mass consistent with $\Xi_c'1\mathrm{P}\big|1/2^-,0\rangle_{\lambda}$ and $\Xi_c'1\mathrm{P}\big|1/2^-,1\rangle_{\lambda}$, which are 2882 MeV and 2867 MeV, respectively. At the same time, as discussed in the previous subsection, $\Xi_c(2882)$ could also be $\Xi_c1\mathrm{P}\big|1/2^-,1\rangle_{\rho}$. Thus, the identification of $\Xi_c(2882)$ requires further determination from its decay properties. The pair of $j$-partner states, $\Xi_c'1\mathrm{P}\big|3/2^-,1\rangle_{\lambda}$ and $\Xi_c'1\mathrm{P}\big|3/2^-,2\rangle_{\lambda}$, have masses of 2930 MeV and 2948 MeV, respectively. These values are reasonably close to the masses of the two experimental resonances $\Xi_c(2923)$ and $\Xi_c(2939)$. Therefore, based on the mass ordering, $\Xi_c(2923)$ and $\Xi_c(2939)$ are tentatively assigned as candidates for $\Xi_c'1\mathrm{P}\big|3/2^-,1\rangle_{\lambda}$ and $\Xi_c'1\mathrm{P}\big|3/2^-,2\rangle_{\lambda}$, respectively. The predicted mass of $\Xi_c'1\mathrm{P}\big|5/2^-,2\rangle_{\lambda}$ is 2969 MeV, and $\Xi_c(2965)$ can be tentatively assigned as the candidate state with this set of quantum numbers with good mass agreement.
It is worth noting that for the $1P$-wave $\lambda$-mode excitations, the predictions for these states from most theoretical studies do not differ drastically. Nevertheless, many experimentally observed masses cluster within this narrow mass window. Small mass differences may therefore lead to different physical interpretations. More precise studies or additional constraints from decay observables are needed to distinguish and confirm these assignments.

For the two $1P$-wave $\rho$-mode $\Xi_c'$ excitations, $\Xi_c'1\mathrm{P}\big|1/2^-,1\rangle_{\rho}$ and $\Xi_c'1\mathrm{P}\big|3/2^-,1\rangle_{\rho}$, our predicted masses are 2946 MeV and 2976 MeV, respectively. Compared with other theoretical results, this result is reasonable. The relevant candidate resonances include $\Xi_c(2939)$ and $\Xi_c(2965)$. Thus, whether they correspond to the $\lambda$-mode or $\rho$-mode $\Xi_c'$ excited states remains undetermined, and further discrimination will be performed by incorporating decay properties in subsequent analysis.

\paragraph*{C. $1D$-wave states}
The $1D$-wave $\Xi_c'$ excited states include six $\lambda$-mode excitations and six $\rho$-mode excitations.
In this work, the masses of the $\lambda$-mode states are about $\sim3150$ MeV, whereas the $\rho$-mode states have a mass of approximately $\sim3250$ MeV. Since these excited states have relatively high masses, only a limited number of experimental resonance candidates are available for matching. The only one that needs to be mentioned is the experimentally observed state $\Xi_c(3123)^+$, which is fairly close to $1D$-wave $\lambda$-mode $\Xi_c'$ excited states.
However, the experimental data for $\Xi_c(3123)^+$ are quite limited. Although it was discovered by the BABAR collaboration in 2008 in the $\Lambda_c^+ \pi^+ K^-$ decay channel~\cite{BaBar:2007zjt}, the Belle collaboration failed to reproduce this signal in the same decay channel in 2014~\cite{Belle:2013htj}. Therefore, a complete determination of the nature of $\Xi_c(3123)$ remains highly challenging.

\subsubsection{Summary of mass spectra}
We investigate the mass spectra of the $1S$-, $1P$-, and $1D$-wave $\lambda/\rho$-mode $\Xi_c$ and $\Xi_c'$ baryons. Based on our mass-spectrum calculations, we attempt to interpret the internal structures of the experimentally observed excited states $\Xi_c(2790)$, $\Xi_c(2815)$, $\Xi_c(2882)$, $\Xi_c(2923)$, $\Xi_c(2939)$, $\Xi_c(2965)$, $\Xi_c(3055)$, $\Xi_c(3080)$ and $\Xi_c(3123)$. Our conclusions are as follows: \romannumeral1) The assignments of $\Xi_c(2790)$, $\Xi_c(2815)$ and $\Xi_c(3055)$ are relatively unambiguous, and correspond to $\Xi_c1\mathrm{P}\big|1/2^-,1\rangle_{\lambda}$, $\Xi_c1\mathrm{P}\big|3/2^-,1\rangle_{\lambda}$, and $\Xi_c1\mathrm{D}\big|3/2^+,2\rangle_{\lambda\lambda}$, respectively. \romannumeral2) $\Xi_c(3080)$ is most likely to be the $\Xi_c1\mathrm{D}\big|5/2^+,2\rangle_{\lambda\lambda}$ state. \romannumeral3) $\Xi_c(2923)$ can be assigned to $\Xi_c'1\mathrm{P}\big|3/2^-,1\rangle_{\lambda}$, and $\Xi_c(2939)$ may be interpreted as either the $\Xi_c'1\mathrm{P}\big|3/2^-,2\rangle_{\lambda}$ state or the $\Xi_c'1\mathrm{P}\big|1/2^-,1\rangle_{\rho}$ state. With relatively limited data on $\Xi_c(2965)$,  it can for now be interpreted as either the $\Xi_c'1\mathrm{P}\big|5/2^-,2\rangle_{\lambda}$ state or the $\Xi_c'1\mathrm{P}\big|3/2^-,1\rangle_{\rho}$ state. \romannumeral4) Since the experimental data for states $\Xi_c(2882)$ and $\Xi_c(3123)$ are quite scarce, their theoretical interpretation remains highly uncertain. $\Xi_c(2882)$ has four possible assignments: $\Xi_c1\mathrm{P}\big|1/2^-,0\rangle_{\rho}$, $\Xi_c1\mathrm{P}\big|1/2^-,1\rangle_{\rho}$, $\Xi_c'1\mathrm{P}\big|1/2^-,0\rangle_{\lambda}$ or $\Xi_c'1\mathrm{P}\big|1/2^-,1\rangle_{\lambda}$. $\Xi_c(3123)$ may be interpreted as $\Xi_c'1\mathrm{D}\big|1/2^+,1\rangle_{\lambda\lambda}$. Naturally, the definitive configuration assignments for all these resonances require further constraints from the subsequent analysis of decay properties.

\begin{table*}[htbp]
	\centering
	\caption{Quark model classifications, calculated masses and decay modes of the $1P$- and $1D$-wave $\Xi_c$ excited states.}
	\label{tab:xic_decay}
	\setlength{\tabcolsep}{5.8pt}
	\fontsize{8.3}{12.5}\selectfont
	\begin{tabular}{c c cp{7.2cm}}
		\noalign{\hrule height 1pt}
		Quark model classifications & Calculated masses (MeV) & Candidate & Decay channels \\
        \hline
  $\Xi_c1\mathrm{P}\big|\tfrac12^-,1\big\rangle_\lambda$ & 2790 &$\Xi_c(2790)$ &$\Xi'^{(*)}_c\pi$ \\
  $\Xi_c1\mathrm{P}\big|\tfrac32^-,1\big\rangle_\lambda$ & 2821&$\Xi_c(2815)$ & $\Xi'^{(*)}_c\pi$ \\
      \hline
  $\Xi_c1\mathrm{P}\big|\tfrac12^-,0\big\rangle_\rho$ & 2834 &$\Xi_c(2882)$ &$ \Xi_c\pi, \;\Xi'^{(*)}_c\pi ,\;\Lambda_c\bar{K}$ \\
  $\Xi_c1\mathrm{P}\big|\tfrac12^-,1\big\rangle_\rho$ & 2861 &$\Xi_c(2882)$& $\Xi_c\pi, \;\Xi'^{(*)}_c\pi ,\;\Lambda_c\bar{K} $ \\
  $\Xi_c1\mathrm{P}\big|\tfrac32^-,1\big\rangle_\rho$ & 3005 & & $\Xi_c\pi, \Xi'^{(*)}_c\pi,\;\Sigma_c K, \Lambda_c\bar{K},\Xi_c(1P_{\lambda/\rho})\pi,  \Xi'_c(1P_\lambda)\pi$ \\
  $\Xi_c1\mathrm{P}\big|\tfrac32^-,2\big\rangle_\rho$ & 3002& & $\Xi_c\pi, \Xi'^{(*)}_c\pi,\;\Sigma_c K, \Lambda_c\bar{K} ,\Xi_c(1P_{\lambda/\rho})\pi,  \Xi'_c(1P_\lambda)\pi$ \\
  $\Xi_c1\mathrm{P}\big|\tfrac52^-,2\big\rangle_\rho$ & 3094 & &$ \Xi_c\pi/\eta, \Xi'^{(*)}_c\pi, \Sigma^{(*)}_c K, \Lambda_c\bar{K},\Xi_c(1P_{\lambda/\rho})\pi,  \Xi'_c(1P_\lambda/\rho)\pi,\Xi_c(2S_{\lambda\lambda})\pi$ \\
		\hline
$\Xi_c1\mathrm{D}\big|\tfrac32^+,2\big\rangle_{\lambda\lambda}$
		& 3051&$\Xi_c(3050)$
		& $ \Xi'^{(*)}_c\pi,\Sigma^{(*)}_c K$ \\[-3pt]
		&&
		& $ \Xi_c(1P_\rho)\pi,\Xi'_c(1P_\lambda)\pi$ \\
		$\Xi_c1\mathrm{D}\big|\tfrac52^+,2\big\rangle_{\lambda\lambda}$
		& 3054&$\Xi_c(3080)$
		& $ \Xi'^{(*)}_c\pi,\Sigma^{(*)}_c K$ \\[-3pt]
		&&
		& $\Xi_c(1P_{\rho})\pi, \Xi'_c(1P_\lambda)\pi$ \\  \hline
  $\Xi_c1\mathrm{D}\big|\tfrac32^+,2\big\rangle_{\rho\rho}$
		& 3249&
		& $ \Xi'^{(*)}_c\pi/\eta, \Sigma^{(*)}_c K, \Omega_c K$ \\[-3pt]
		&&
		& $\Xi_c(1P_{\rho})\pi, \Xi'_c(1P_\lambda)\pi, \Sigma_c(1P_\lambda) K$ \\[-3pt]
		&&
		& $ \Xi'_c(2S_{\lambda\lambda/\rho\rho})\pi, \Xi_c(1D_{\lambda\lambda/\rho\rho})\pi$ \\
  $\Xi_c1\mathrm{D}\big|\tfrac52^+,2\big\rangle_{\rho\rho}$
		& 3288&
		& $\Xi'^{(*)}_c\pi/\eta,\Omega^{(*)}_c K,\Sigma^{(*)}_c K$ \\[-3pt]
		&&
		& $\Xi_c(1P_{\rho})\pi, \Lambda_c(1P_{\rho})\bar{K}, \Xi'_c(1P_\lambda)\pi,\Sigma_c(1P_\lambda) K$ \\[-3pt]
		&&
		& $ \Xi'_c(2S_{\lambda\lambda/\rho\rho})\pi, \Xi_c'(1D_{\lambda\lambda/\rho\rho})\pi$ \\
		\noalign{\hrule height 1pt}
	\end{tabular}
\end{table*}

\subsection{Strong Decay}

In the previous subsection, we discussed the mass spectra of $\Xi_c$ and $\Xi_c'$ states. Based on the obtained mass spectra, we attempt to interpret the controversial excited states and clarify their inner structures. However, because the (2900$\sim$3100) MeV mass region contains numerous states, an interpretation of the disputed states based solely on mass remains insufficient. Next, we further use the realistic wave functions obtained from the calculated spectra to study the strong decay properties within the QPC model in the $j-j$ coupling scheme, so as to further determine the internal structures of the excited states. For ease of calculation, a harmonic oscillator wave function is adopted instead of the realistic wave function. The main conclusions remain qualitatively valid.

\subsubsection{Strong decay of the $\Xi_c$ resonances}

In the framework of the QPC model, we systematically investigate the Okubo–Zweig–Iizuka(OZI)-allowed two body strong decay properties of the $1P$- and $1D$-wave $\Xi_c$ states within the $j-j$ coupling scheme. The corresponding quark model classification and possible strong decay modes are collected in Table~\ref{tab:xic_decay}.

\paragraph*{A. $1P$-wave states}
For the two $1P$-wave $\lambda$-mode $\Xi_c$ excited baryons,
$\Xi_c1P\big|J^P=1/2^-,1\big\rangle_\lambda$ and $\Xi_c1P\big|J^P=3/2^-,1\big\rangle_\lambda$,
we argue that the experimentally observed states
$\Xi_c(2790)$ and $\Xi_c(2815)$  can serve as the respective candidates based the our mass spectrum results, and this assignments have been adopted in numerous references. Therefore, we directly fix the masses of $\Xi_c1P\big|J^P=1/2^-,1\big\rangle_\lambda$ and $\Xi_c1P\big|J^P=3/2^-,1\big\rangle_\lambda$ to the physical masses, 2792 MeV and 2817 MeV, and analysis their decay properties. The obtained partial and total decay widths are collected in Table~\ref{tab:xic_1Plambda_decay}. Based on our numerical results, both $\Xi_c1P\big|J^P=1/2^-,1\big\rangle_\lambda$ and $\Xi_c1P\big|J^P=3/2^-,1\big\rangle_\lambda$ are narrow states with decay widths of a few MeV. However, their main decay channels are different.

For the $\Xi_c1P\big|J^P=1/2^-,1\big\rangle_\lambda$ state, the predicted total decay width is
\[
\Gamma_{\text{Total}} \simeq 9.51\,\text{MeV},
\]
 and it is almost entirely dominated by decay channel $\Xi_c'\pi$. These decay properties are highly consistent with those of the experimentally observed state $\Xi_c(2790)$. It should be noted that the new decay channel $\Lambda_cK$ of $\Xi_c(2790)$, observed by the LHCb Collaboration~\cite{LHCb:2022vns} in 2023, is forbidden by the spin selection rules in our calculations and thus relatively narrow. However, its experimental significance is only 3.7$\sigma$. Therefore, $\Xi_c(2790)$ can still be reasonably interpreted as the candidate of the $\Xi_c1P\big|J^P=1/2^-,1\big\rangle_\lambda$ state to some extent.

\begin{table}[htb]
	\centering
	\caption{The strong decay properties of the $1P$-wave $\lambda$-mode $\Xi_c$ states, which are taken as $\Xi_c(2790)$ and $\Xi_c(2815)$, respectively. $\Gamma_{\text{Total}}$ represents the total decay width, and $\Gamma_{\text{Expt.}}$ denotes the experimental value. The unit is MeV.}
	\label{tab:xic_1Plambda_decay}
	\setlength{\tabcolsep}{2.8pt}
	\begin{tabular}{lcc}
		\hline\hline
		& \underline{$\Xi_c\big|J^P=\tfrac{1}{2}^-,1\big\rangle_\lambda$}
		& \underline{$\Xi_c\big|J^P=\tfrac{3}{2}^-,1\big\rangle_\lambda$} \\
		Decay width & $\Xi_c(2790)$ & $\Xi_c(2815)$ \\
		\hline
		$\Gamma[\Xi_c'\pi]$ & $9.51$& $0.09$  \\
		$\Gamma[\Xi_c'^*\pi]$ & $0.00$ & $4.86$ \\
		\hline
		$\Gamma_{\text{Total}}$ & $9.51$& $4.95$  \\
		\hline
		$\Gamma_{\text{Expt.}}$ & $8.9\pm0.6\pm0.8$& $2.15\pm0.15$  \\
		\hline\hline
	\end{tabular}
\end{table}

The total decay width of $\Xi_c1P\big|J^P=3/2^-,1\big\rangle_\lambda$ is predicted to be
\[
\Gamma_{\text{Total}} \simeq 4.95\,\text{MeV},
\]
which is approximately twice the experimental total width of $\Xi_c(2815)$. The dominant channel is $\Xi_c'^*\pi$ with a branching ratio
\[
\frac{\Gamma\big[\Xi_c1P\big|J^P=3/2^-,1\big\rangle_\lambda \to \Xi_c'^*\pi\big]}{\Gamma_{\text{Total}}} \simeq 98\%.
\]
The large branching ratio indicates that $\Xi_c1P\big|J^P=3/2^-,1\big\rangle_\lambda$ is likely to be observed in the $\Xi_c\pi\pi$ final state via the decay chain $\Xi_c1P\big|J^P=3/2^-,1\big\rangle_\lambda\rightarrow\Xi_c^*\pi\rightarrow\Xi_c\pi\pi$.
The other decay channel $\Xi_c'\pi$ has a relatively small branching ratio, which is about
\[
\frac{\Gamma\big[\Xi_c1P\big|J^P=3/2^-,1\big\rangle_\lambda \to \Xi_c'\pi\big]}{\Gamma_{\text{Total}}} \simeq 2\%.
\]
Combining the decay properties of the state $\Xi_c(2815)$ being observed in the decay channels $\Xi_c'\pi$, $\Xi_c'^*\pi$, and $\Xi_c\pi\pi$, it is reasonable to interpret the state $\Xi_c(2815)$ as the state $\Xi_c1P\big|J^P=3/2^-,1\big\rangle_\lambda$.

Therefore, based on our theoretical results from mass spectrum and decay calculations, $\Xi_c1P\big|J^P=1/2^-,1\big\rangle_\lambda$ and $\Xi_c1P\big|J^P=3/2^-,1\big\rangle_\lambda$ are good candidates for the experimentally observed states $\Xi_c(2790)$ and $\Xi_c(2815)$, respectively.

\begin{table*}[tp]
	\centering
	\caption{The strong decay properties of the $1P$-wave $\rho$-mode $\Xi_c$ states. The unit is MeV.}
	\label{tab:xic_1Prho_decay1}
	\setlength{\tabcolsep}{2.8pt}
	\hspace{-15mm} 
	\fontsize{8.5}{8.5}\selectfont
	\begin{tabular}{l *{11}{c}}
		\hline\hline
		& \multicolumn{2}{c}{$\Xi_c1P\big|\tfrac{1}{2}^-,0\big\rangle_\rho$}
		& \multicolumn{2}{c}{$\Xi_c1P\big|\tfrac{1}{2}^-,1\big\rangle_\rho$}
		& \multicolumn{2}{c}{$\Xi_c1P\big|\tfrac{3}{2}^-,1\big\rangle_\rho$}
		& \multicolumn{2}{c}{$\Xi_c1P\big|\tfrac{3}{2}^-,2\big\rangle_\rho$}
		& \multicolumn{2}{c}{$\Xi_c1P\big|\tfrac{5}{2}^-,2\big\rangle_\rho$} \\[-4pt]
		& \multicolumn{2}{c}{\rule{2.6cm}{0.4pt}}
		& \multicolumn{2}{c}{\rule{2.6cm}{0.4pt}}
		& \multicolumn{2}{c}{\rule{1.6cm}{0.4pt}}
		& \multicolumn{2}{c}{\rule{1.6cm}{0.4pt}}
		& \multicolumn{2}{c}{\rule{1.6cm}{0.4pt}} \\[-2pt]
		Decay Width & $2834$ & $\Xi_c(2882)$ & $2861$ &$\Xi_c(2882)$ & $3005$ & & $3002$ & & $3094$&   \\
		\hline
		$\Gamma[\Xi_c\pi/\eta]$
		& $13.27^*$/-- & $13.28^*$/-- & $8.90^*$/-- & $8.70$ & $0.01$/-- &  & $6.20$/-- &    & $14.27/0.49$ &    \\
		$\Gamma[\Xi_c'\pi]$
		& $4.69^*$ & $5.59$ & $18.92^*$ & $20.23$ & $1.25$ &   & $2.04$ &  & $2.92$ &  \\
		$\Gamma[\Xi_c'^*\pi]$
		& $0.00$ & $0.01$ & $0.01$ & $0.02$ & $33.28^*$ & & $1.11$ &  & $6.22$ &   \\
		$\Gamma[\Lambda_c\bar{K}]$
		& $11.07^*$ & $11.94$ & $8.78^*$ & $8.64$ & $0.00$ &  & $5.62$ & & $12.56$ &   \\
		$\Gamma[\Sigma_c K]$
		& -- & -- & -- & --  & $0.50$ &  & $0.75$ & & $4.10$ &  \\
		$\Gamma[\Sigma_c^* K]$
		& --& --  & -- & --  & -- & & -- &  & $4.20$ & \\
		$\Gamma\big[\Xi_c1P\big|\tfrac{1}{2}^-,0/1\big\rangle_\rho \pi\big]$
		& --/-- & --/-- & --/-- & --/-- & $0.07/0.01$ &   & $0.04/0.00$ &      & $0.00/0.00$ &  \\
		$\Gamma\big[\Xi_c1P\big|\tfrac{1}{2}^-/\tfrac{3}{2}^-,1\big\rangle_\lambda \pi\big]$
		& --/-- &--/-- & --/-- &--/-- & $0.00/0.00$ &   & $0.06/0.00$ &       & $0.03/0.84$ &   \\
		$\Gamma\big[\Xi_c'1P\big|\tfrac{1}{2}^-,1\big\rangle_\rho \pi\big]$
		& -- &-- & -- & -- & -- & & -- &       & $0.00$ &  \\
		$\Gamma\big[\Xi_c'1P\big|\tfrac{1}{2}^-,0/1\big\rangle_\lambda \pi\big]$
		& --/-- &--/-- & --/-- & --/-- & --/0.00& & --/0.00 && 0.00/$0.00$ &     \\
		$\Gamma\big[\Xi_c'1P\big|\tfrac{3}{2}^-,1/2\big\rangle_\lambda \pi\big]$
		& --/-- &--/-- & --/-- & --/-- & --/-- && --/-- &      & $0.03/0.00$ &   \\
		$\Gamma\big[\Xi_c\ 2S\big|\tfrac{1}{2}^+,1\big\rangle_{\lambda} \pi \big]$
		& -- & -- & -- & -- & -- &   & -- &    & $0.00$ & \\	
		\hline 
		$\Gamma_{\text{Total}}$
		&$29.03$ & $30.82$ & $36.61$& $37.59$  & $35.12$ &   & $15.82$ &    & $45.66$ &  \\
		\hline
		$\Gamma_{\text{Expt.}}$
		&\multicolumn{4}{c}{$12.4\pm5.3\pm5.8$}  & & & &  & & & \\
		\hline\hline
	\end{tabular}
\end{table*}

For the five $1P$-wave $\rho$-mode excited $\Xi_c$ baryons, $\Xi_c\big|J^P=1/2^-,0\big\rangle_\rho$, $\Xi_c\big|J^P=1/2^-,1\big\rangle_\rho$, $\Xi_c\big|J^P=3/2^-,1\big\rangle_\rho$, $\Xi_c\big|J^P=3/2^-,2\big\rangle_\rho$,
and $\Xi_c\big|J^P=5/2^-,2\big\rangle_\rho$, we first fix their masses at the calculated mass-spectrum values and then study their decay properties, as shown in Table~\ref{tab:xic_1Prho_decay1}.

From the table, $\Xi_c1P\big|J^P=1/2^-,0\big\rangle_\rho$ and $\Xi_c1P\big|J^P=1/2^-,1\big\rangle_\rho$ have highly similar decay properties. They are moderate states with decay widths of several tens of MeV and mainly decay into the final states $\Xi_c\pi$, $\Xi_c'\pi$, and $\Lambda_c \bar{K}$. In our mass-spectrum analysis, $\Xi_c(2882)$ is regarded as a candidate for $\Xi_c1P\big|J^P=1/2^-,0\big\rangle_\rho$ or $\Xi_c1P\big|J^P=1/2^-,1\big\rangle_\rho$. Hence, we further fix the masses of $\Xi_c1P\big|J^P=1/2^-,0\big\rangle_\rho$ and $\Xi_c1P\big|J^P=1/2^-,1\big\rangle_\rho$ on the physical mass 2882 MeV, and calculate their decay properties(shown in Table~\ref{tab:xic_1Prho_decay1}).

The total decay width of $\Xi_c1P\big|J^P=1/2^-,0\big\rangle_\rho$ is
\[
\Gamma_{\text{Total}} \simeq 30.82\,\text{MeV},
\]
which is marginally larger than the experimental upper limit on the total width of $\Xi_c(2882)$. The dominant decay channels are $\Xi_c\pi$ and $\Lambda_c\bar{K}$. The corresponding predicted branching fractions are
\[
\frac{\Gamma\big[\Xi_c1P\big|J^P=1/2^-,0\big\rangle_\rho \to \Xi_c\pi\big]}{\Gamma_{\text{Total}}} \simeq 43\%,
\]
and
\[
\frac{\Gamma\big[\Xi_c1P\big|J^P=1/2^-,0\big\rangle_\rho \to \Lambda_c\bar{K}\big]}{\Gamma_{\text{Total}}} \simeq 39\%,
\]
respectively. The large branching ratio of decay channel $\Lambda_c\bar{K}$ is also in agreement with the experimental fact that
$\Xi_c(2882)$ was observed in decay channel $\Lambda_c\bar{K}$. Therefore, taking into account both the mass and decay properties, $\Xi_c(2882)$ is very likely to be a candidate for $\Xi_c1P\big|J^P=1/2^-,0\big\rangle_\rho$.

The total decay width of state $\Xi_c1P\big|J^P=1/2^-,1\big\rangle_\rho$ is slightly larger than that of state $\Xi_c1P\big|J^P=1/2^-,0\big\rangle_\rho$, about
\[
\Gamma_{\text{Total}} \simeq 37.59\,\text{MeV}.
\]
Its dominant decay is into the final state $\Xi_c'\pi$, with a branching ratio of
\[
\frac{\Gamma\big[\Xi_c1P\big|J^P=1/2^-,1\big\rangle_\rho \to \Xi_c'\pi\big]}{\Gamma_{\text{Total}}} \simeq 54\%.
\]
At the same time, decay channels $\Xi_c\pi$ and $\Lambda_c\bar{K}$ also have non-negligible branching ratios, which are
\[
\frac{\Gamma\big[\Xi_c1P\big|J^P=1/2^-,1\big\rangle_\rho \to \Xi_c\pi\big]}{\Gamma_{\text{Total}}} \simeq 23\%,
\]
and
\[
\frac{\Gamma\big[\Xi_c1P\big|J^P=1/2^-,1\big\rangle_\rho \to \Lambda_c\bar{K}\big]}{\Gamma_{\text{Total}}} \simeq 23\%,
\]
respectively. Considering both theoretical and experimental uncertainties, the possibility that $\Xi_c(2882)$ is a candidate for the quark configuration $\Xi_c1P\big|J^P=1/2^-,1\big\rangle_\rho$ cannot be excluded.

\begin{figure*}[]
  \centering
  \includegraphics[width=0.85\textwidth]{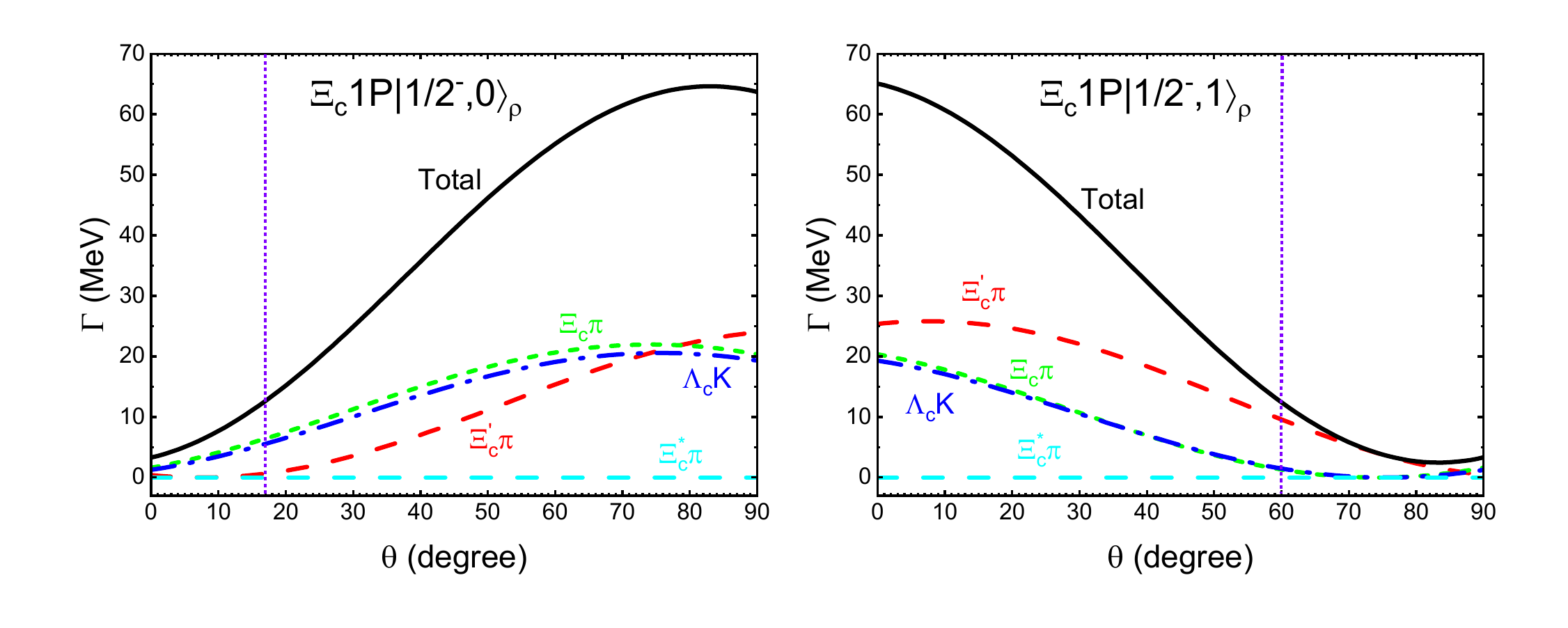}
  \caption{Partial and total strong decay widths of $\Xi_c1P\big|J^P=1/2^-,0\big\rangle_\rho$ and $\Xi_c1P\big|J^P=1/2^-,1\big\rangle_\rho$ as functions of the mixing angle $\theta$.}
  \label{Fig:Xic1Pr10and1_theta}
\end{figure*}

Additionally, we know that the states in the $j-j$ coupling scheme are the linear combination of the configurations in the $L-S$ coupling scheme. That means $\Xi_c1P\big|J^P=1/2^-,0\big\rangle_\rho$ and $\Xi_c1P\big|J^P=1/2^-,1\big\rangle_\rho$ in the $j-j$ coupling scheme contain a mixing angle $\theta$. Considering the heavy quark symmetry being not strictly true and slightly breaking in the single-charm strange system, the mixing angle $\theta$ will fluctuate around the center value($\theta\simeq35^{\circ}$). To investigate this effect, we plot the strong decay properties of $\Xi_c1P\big|J^P=1/2^-,0\big\rangle_\rho$ and $\Xi_c1P\big|J^P=1/2^-,1\big\rangle_\rho$ within the QPC model as functions of the mixing angle varying in region of $\theta=(0^{\circ}\sim90^{\circ})$ in Fig.~\ref{Fig:Xic1Pr10and1_theta}.
As shown in the figure, when the mixing angle $\theta\simeq17^{\circ}$, the total decay width of $\Xi_c1P\big|J^P=1/2^-,0\big\rangle_\rho$ is $\Gamma\simeq12.67$ MeV, which is consistent with that of the the observed state $\Xi_c(2882)$. Furthermore, the branching fractions of the dominant decay channels $\Xi_c\pi$ and $\Lambda_c\bar{K}$ are 51$\%$ and 44$\%$, respectively. For the state $\Xi_c1P\big|J^P=1/2^-,1\big\rangle_\rho$, when its total decay width is consistent with the central value of the $\Xi_c(2882)$ width, the corresponding mixing angle is approximately $\theta\simeq60^{\circ}$. In this scenario, the branching ratio of the dominant decay channel $\Xi_c'\pi$ increases to about 77$\%$, while those of decay channels $\Xi_c\pi$ and $\Lambda_c\bar{K}$ decrease to 11$\%$ and 12$\%$, respectively. Therefore, if $\Xi_c(2882)$ is indeed state $\Xi_c1P\big|J^P=1/2^-,1\big\rangle_\rho$, decay channel $\Xi_c'\pi$ would be a very promising channel to observed experimentally.

The state $\Xi_c1P\big|J^P=3/2^-,1\big\rangle_\rho$ has a decay width of approximately $\Gamma\sim35$ MeV, and its dominant decay channel is $\Xi_c'^*\pi$. Given its relatively narrow width, this state is very likely to be observed experimentally in the $\Xi_c\pi\pi$ final state via the decay chain $\Xi_c1P\big|J^P=3/2^-,1\big\rangle_\rho\rightarrow \Xi_c'^*\pi\rightarrow \Xi_c\pi\pi$.

The state $\Xi_c1P\big|J^P=3/2^-,2\big\rangle_\rho$ is likely a relatively narrow state, with a width of approximately $\Gamma\sim16$ MeV. Its dominant decay channels are $\Xi_c\pi$ and $\Lambda_c\bar{K}$, with branching ratios of about 39$\%$ and 35$\%$, respectively. We note that the decay properties of this state agree well with those of $\Xi_c(2882)$. However, since its mass is over 100 MeV higher, we do not currently consider it a candidate for $\Xi_c(2882)$.

The width of state $\Xi_c\big|J^P=5/2^-,2\big\rangle_\rho$ is slightly larger, approximately $\Gamma\sim46$ MeV, and it mainly decays into channels $\Xi_c\pi$ and $\Lambda_c\bar{K}$. This state may be observed experimentally in its two main decay channels.

\paragraph*{B. $1D$-wave states}

\begin{table}[htb]
	\centering
	\caption{The strong decay properties of the $1D$-wave $\lambda$-mode $\Xi_c$ states. The unit is MeV.}
	\label{tab:xic_1Dlambdalambda_decay}
	\setlength{\tabcolsep}{4.6pt}
	\fontsize{7.5}{7.5}\selectfont
	\begin{tabular}{l *{4}{c}}
		\hline\hline
		& \multicolumn{2}{c}{$\Xi_c1D\big|J^P=\tfrac{3}{2}^+,2\big\rangle_{\lambda\lambda}$}
		& \multicolumn{2}{c}{$\Xi_c1D\big|J^P=\tfrac{5}{2}^+,2\big\rangle_{\lambda\lambda}$} \\[-3pt]
		& \multicolumn{2}{c}{\rule{2.3cm}{0.4pt}}
		& \multicolumn{2}{c}{\rule{2.3cm}{0.4pt}} \\[-3pt]
		Decay Width & $3051$ & $\Xi_c(3055)$ & $3054$ & $\Xi_c(3080)$ \\
		\hline
		$\Gamma[\Xi_c'\pi]$
		& $2.15$ & $2.17$ & $0.19$ & $0.27$ \\
		$\Gamma[\Xi_c'^*\pi]$
		& $0.46$ & $0.48$ & $2.83$ & $3.16$ \\
		$\Gamma[\Sigma_c/\Sigma_c^* K]$
		& $3.89/0.24$ & $4.04/0.28$ & $0.08/2.07$ & $0.17/3.68$  \\
		$\Gamma\big[\Xi_c\ 1P\big|J^P=\tfrac{1}{2}^-,0/1\big\rangle_\rho \pi\big]$
		& $0.02/0.00$ & $0.02/0.00$ & $0.02/0.00$ & $0.03/0.00$ \\
		$\Gamma\big[\Xi_c'\ 1P\big|J^P=\tfrac{1}{2}^-,0/1\big\rangle_\lambda \pi\big]$
		& $0.00/0.00$ & $0.00/0.00$ & $0.00/0.00$ & $0.01/0.01$ \\	
		\hline
		$\Gamma_{\text{Total}}$
		& $6.76$ & $6.99$ & $5.19$ & $7.33$ \\
		\hline
		$\Gamma_{\text{Expt.}}$
		& \multicolumn{2}{c}{$8.0\pm0.8$} & \multicolumn{2}{c}{$3.8\pm0.9$} \\
		\hline\hline
	\end{tabular}
\end{table}

For the two $1D$-wave $\lambda$-excited $\Xi_c$ baryons $\Xi_c1D\big|J^P=3/2^+,2\big\rangle_{\lambda\lambda}$ and $\Xi_c1D\big|J^P=5/2^+,2\big\rangle_{\lambda\lambda}$, similarly, we first fix their masses to the values obtained from the mass-spectrum calculation and then obtain their strong-decay properties, as shown in Table~\ref{tab:xic_1Dlambdalambda_decay}. It is shown that the two states are both narrow states with total decay widths only several MeV. The dominant decay channels of $\Xi_c1D\big|J^P=3/2^+,2\big\rangle_{\lambda\lambda}$ are $\Sigma_cK$ and $\Xi_c'\pi$, while those of $\Xi_c1D\big|J^P=5/2^+,2\big\rangle_{\lambda\lambda}$ are $\Sigma_c^*K$ and $\Xi_c'^*\pi$.

Based on the preceding analysis of the mass spectrum, we find that states $\Xi_c1D\big|J^P=3/2^+,2\big\rangle_{\lambda\lambda}$ and $\Xi_c1D\big|J^P=5/2^+,2\big\rangle_{\lambda\lambda}$ are likely candidate states for the experimentally observed states $\Xi_c(3055)$ and $\Xi_c(3080)$. We therefore further fix the masses of $\Xi_c1D\big|J^P=3/2^+,2\big\rangle_{\lambda\lambda}$ and $\Xi_c1D\big|J^P=5/2^+,2\big\rangle_{\lambda\lambda}$ to their physical masses of 3055 MeV and 3077 MeV, respectively, and discuss their strong-decay properties, which are also listed in Table~\ref{tab:xic_1Dlambdalambda_decay}.

We can obtain that the total decay width of $\Xi_c1D\big|J^P=3/2^+,2\big\rangle_{\lambda\lambda}$ is
\[
\Gamma_{\text{Total}} \simeq 6.99\,\text{MeV}.
\]
Within the error range, this value is consistent with the total width of the experimentally observed state $\Xi_c(3055)$. The branching ratios of the two dominant decay channels $\Sigma_cK$ and $\Xi_c'\pi$ are predicted to be
\[
\frac{\Gamma\big[\Xi_c1D\big|J^P=3/2^+,2\big\rangle_{\lambda\lambda} \to \Sigma_cK\big]}{\Gamma_{\text{Total}}} \simeq 58\%,
\]
and
\[
\frac{\Gamma\big[\Xi_c1D\big|J^P=3/2^+,2\big\rangle_{\lambda\lambda} \to \Xi_c'\pi\big]}{\Gamma_{\text{Total}}} \simeq 31\%,
\]
respectively. Hence, this state is highly likely to be observed experimentally in the $\Lambda_c\pi K$ final state via the decay chain $\Xi_c1D\big|J^P=3/2^+,2\big\rangle_{\lambda\lambda}\rightarrow \Sigma_cK\rightarrow\Lambda_c\pi K$.  Meanwhile, if the state $\Xi_c(3055)$ is indeed interpretable as the quark configuration $\Xi_c1D\big|J^P=3/2^+,2\big\rangle_{\lambda\lambda}$, then, apart from the decay channel $\Sigma_cK$, the decay channel $\Xi_c'\pi$ can also be used as an alternative decay channel for the experimental study of state  $\Xi_c(3055)$.

For the states $\Xi_c1D\big|J^P=5/2^+,2\big\rangle_{\lambda\lambda}$, the total decay width is
\[
\Gamma_{\text{Total}} \simeq 7.33\,\text{MeV},
\]
which is approximately twice the total width of the observed state $\Xi_c(3080)$.
The dominant decay channels are $\Xi_c'^*\pi$ and $\Sigma_c^* K$, with branching ratios
\[
\frac{\Gamma\big[\Xi_c1D\big|J^P=5/2^+,2\big\rangle_{\lambda\lambda} \to \Xi_c'^*\pi\big]}{\Gamma_{\text{Total}}} \simeq43\%,
\]
and
\[
\frac{\Gamma\big[\Xi_c1D\big|J^P=5/2^+,2\big\rangle_{\lambda\lambda} \to \Sigma_c^* K\big]}{\Gamma_{\text{Total}}} \simeq 50\%.
\]
Hence,  $\Xi_c1D\big|J^P=5/2^+,2\big\rangle_{\lambda\lambda}$ may be observed experimentally in the final states $\Xi_c\pi\pi$ and $\Lambda_c\pi K$ via the decay chains $\Xi_c1D\big|J^P=5/2^+,2\big\rangle_{\lambda\lambda}\rightarrow\Xi_c'^*\pi\rightarrow\Xi_c\pi\pi$ and $\Xi_c1D\big|J^P=5/2^+,2\big\rangle_{\lambda\lambda}\rightarrow \Sigma_c^*K\rightarrow\Lambda_c\pi K$. These decay properties are consistent with those of the observed state $\Xi_c(3080)$. Therefore, combining our theoretical predictions of mass, decay channels and decay width,
$\Xi_c(3080)$ can be assigned as the quantum state $\Xi_c1D\big|J^P=5/2^+,2\big\rangle_{\lambda\lambda}$.

In summary, based on our theoretical results from mass spectrum and decay calculations,
$\Xi_c(3055)$ and $\Xi_c(3080)$ are good candidates for $\Xi_c1D\big|J^P=3/2^+,2\big\rangle_{\lambda\lambda}$ and $\Xi_c1D\big|J^P=5/2^+,2\big\rangle_{\lambda\lambda}$, respectively.

\begin{table}[htb]
	\centering
	\caption{The strong decay properties of the $1D$-wave $\rho$-mode $\Xi_c$ states. The unit is MeV.}
	\label{tab:xic_1Drhorho_decay}
	\setlength{\tabcolsep}{4.6pt}
	\fontsize{8.5}{8.5}\selectfont
	\begin{tabular}{l *{4}{c}}
		\hline\hline
		& \multicolumn{2}{c}{$\Xi_c\big|J^P=\tfrac{3}{2}^+,2\big\rangle_{\rho\rho}$}
		& \multicolumn{2}{c}{$\Xi_c\big|J^P=\tfrac{5}{2}^+,2\big\rangle_{\rho\rho}$}
		\\
		\cmidrule(lr){2-3} \cmidrule(lr){4-5}
		Decay Width& $M=3249$ & & $M=3288$ & \\
		\hline
		
		$\Gamma[\Xi_c'\pi/\eta]$
		& $2.15/1.21$ & & $2.27/0.12$ & \\
		$\Gamma[\Xi_c'^*\pi/\eta]$
		& $2.46/0.13$ & & $6.66/2.06$ & \\
		
		$\Gamma[\Omega_c/\Omega_c^* K]$
		& $0.72$/-- & & $0.02/0.57$ & \\
		$\Gamma[\Sigma_c/\Sigma_c^* K]$
		& $7.04/5.46$ & & $5.31/19.77$ &  \\
		
		$\Gamma\big[\Xi_c\ 1P\big|J^P=\tfrac{1}{2}^-,0/1\big\rangle_\rho \pi\big]$
		& $0.46/0.08$ & & $1.12/0.00$ & \\
		$\Gamma\big[\Xi_c\ 1P\big|J^P=\tfrac{3}{2}^-,1/2\big\rangle_\rho \pi\big]$
		& $0.60/9.69$ & & $1.02/0.83$ &  \\
		$\Gamma\big[\Xi_c\ 1P\big|J^P=\tfrac{5}{2}^-,2\big\rangle_\rho \pi\big]$
		& $0.15$ & & $7.88$ &  \\
		$\Gamma\big[\Lambda_c\ 1P\big|J^P=\tfrac{1}{2}^-,0\big\rangle_\rho \bar{K}\big]$
		& -- & & $1.02$ & \\
		
		$\Gamma\big[\Xi_c'\ 1P\big|J^P=\tfrac{1}{2}^-,0/1\big\rangle_\lambda \pi\big]$
		& $0.00/0.21$ & & $0.02/0.17$ &  \\	
		$\Gamma\big[\Xi_c'\ 1P\big|J^P=\tfrac{3}{2}^-,1/2\big\rangle_\lambda \pi\big]$
		& $0.08/0.06$ & & $0.28/0.01$ &  \\	
		$\Gamma\big[\Xi_c'\ 1P\big|J^P=\tfrac{5}{2}^-,2\big\rangle_\lambda \pi\big]$
		& $0.00$ & & $0.69$ &  \\	
		$\Gamma\big[\Sigma_c\ 1P\big|J^P=\tfrac{1}{2}^-,1\big\rangle_\lambda \ K\big]$
		& $0.07$ & & $0.17$ & \\	
		$\Gamma\big[\Sigma_c\ 1P\big|J^P=\tfrac{3}{2}^-,2\big\rangle_\lambda \ K\big]$
		& -- & & $0.00$ & \\	
		$\Gamma\big[\Sigma_c\ 1P\big|J^P=\tfrac{5}{2}^-,2\big\rangle_\lambda \ K \big]$
		& -- & & $0.40$ &  \\	
		
		$\Gamma\big[\Xi_c'\ 2S\big|J^P=\tfrac{1}{2}^+/\tfrac{3}{2}^+,1\big\rangle_{\rho\rho} \pi \big]$
		& $0.00$/-- & & $0.00/0.00$ &  \\	
		$\Gamma\big[\Xi_c'\ 2S\big|J^P=\tfrac{1}{2}^+/\tfrac{3}{2}^+,1\big\rangle_{\lambda\lambda} \pi \big]$
		& $0.00/0.00$ & & $0.00/0.00$ & \\	
		
		$\Gamma\big[\Xi_c'\ 1D\big|J^P=\tfrac{1}{2}^+,1\big\rangle_{\rho\rho} \pi \big]$
		& $0.04$ & & $0.00$ & \\	
		$\Gamma\big[\Xi_c'\ 1D\big|J^P=\tfrac{1}{2}^+,1\big\rangle_{\lambda\lambda} \pi \big]$
		& -- & & $0.00$ & \\	
		\hline
		$\Gamma_{\text{Total}}$
		& $30.61$ & & $50.39$ & \\
		\hline\hline
	\end{tabular}
\end{table}

The decay properties of the two $1D$-wave $\rho$-mode $\Xi_c$ resonances $\Xi_c1D\big|J^P=3/2^+,2\big\rangle_{\rho\rho}$ and $\Xi_c1D\big|J^P=5/2^+,2\big\rangle_{\rho\rho}$ are collected in Table~\ref{tab:xic_1Drhorho_decay}. From the table, it is seen that these two states are likely to be moderately broad, with total widths of approximately several tens of MeV. They share the same dominant decay channels, which are $\Sigma_c^{(*)}K$ and $\Xi_c'^*\pi$. Therefore, in the mass range (3200$\sim$3300 MeV), these two states are highly likely to be observed experimentally in the final states $\Xi_c\pi\pi$ and $\Lambda_c\pi K$ via the decay chains $\Xi_c1D\big|J^P=3/2^+(5/2^+),2\big\rangle_{\rho\rho}\rightarrow\Xi_c'^*\pi\rightarrow\Xi_c\pi\pi$ and $\Xi_c1D\big|J^P=3/2^+(5/2^+),2\big\rangle_{\rho\rho}\rightarrow \Sigma_c^{(*)}K\rightarrow\Lambda_c\pi K$.

\begin{table*}[htbp]
	\centering
	\caption{Quark model classifications, calculated masses and decay modes of the $1P$- and $1D$-wave $\Xi_c'$ excited states.}
	\setlength{\tabcolsep}{3.8pt}
	\fontsize{7.3}{9.5}\selectfont
	\begin{tabular}{l c c p{8.2cm}}
		\hline
		Quark model classifications & Calculated mass (MeV) &Candidate& Decay channels \\
		\hline
		$\Xi_c'1P\left|\frac{1}{2}^-,0\right\rangle_\lambda$ & 2882& &$\Xi_c\pi, \Xi'^{(*)}_c\pi,\Lambda_c\bar{K}$  \\
		$\Xi_c'1P\left|\frac{1}{2}^-,1\right\rangle_\lambda$ & 2867 && $\Xi_c\pi, \Xi'^{(*)}_c\pi,\Lambda_c\bar{K}$ \\
		$\Xi_c'1P\left|\frac{3}{2}^-,1\right\rangle_\lambda$ & 2930 &$\Xi_c(2923)$& $\Xi_c\pi, \Xi'^{(*)}_c\pi,\Lambda_c\bar{K}$ \\
		$\Xi_c'1P\left|\frac{3}{2}^-,2\right\rangle_\lambda$ & 2948 &$\Xi_c(2939)$& $\Xi_c\pi, \Xi'^{(*)}_c\pi,\Lambda_c\bar{K},\Xi_c(1P_{\lambda})\pi$ \\
		$\Xi_c'1P\left|\frac{5}{2}^-,2\right\rangle_\lambda$ & 2969 &$\Xi_c(2965)$& $\Xi_c\pi, \Xi'^{(*)}_c\pi,\Sigma_c K,\Lambda_c\bar{K},\Xi_c(1P_{\lambda})\pi$ \\
		\hline
		$\Xi_c'1P\left|\frac{1}{2}^-,1\right\rangle_\rho$ & 2946 &&  $ \Xi'^{(*)}_c\pi$\\
		$\Xi_c'1P\left|\frac{3}{2}^-,1\right\rangle_\rho$ & 2976 && $ \Xi'^{(*)}_c\pi,\Sigma_c K$ \\
        \hline
        $\Xi_c'1D\left|\frac{1}{2}^+,1\right\rangle_{\lambda\lambda}$
		& 3153&$\Xi_c(3123)$
		& $\Xi_c\pi/\eta, \Xi'^{(*)}_c\pi, \Xi'_c\eta, \Sigma^{(*)}_c K, \Lambda_c\bar{K},\Lambda/\Sigma D,$ \\[-3pt]
		&&
		& $\Xi_c(1P_{\lambda/\rho})\pi, \Xi'_c(1P_\lambda/\rho)\pi,$ \\[-3pt]
		&&
		& $\Xi_c(2S_{\lambda\lambda})\pi, \Xi'_c(2S_{\lambda\lambda})\pi$ \\
		$\Xi_c'1D\left|\frac{3}{2}^+,1\right\rangle_{\lambda\lambda}$
		& 3181&
		& $\Xi_c\pi/\eta, \Xi'^{(*)}_c\pi, \Xi'_c\eta, \Sigma^{(*)}_c K, \Lambda_c\bar{K},\Lambda/\Sigma D,$ \\[-3pt]
		&&
		& $\Xi_c(1P_{\lambda/\rho})\pi, \Xi'_c(1P_\lambda/\rho)\pi,$ \\[-3pt]
		&&
		& $\Xi_c(2S_{\lambda\lambda})\pi, \Xi'_c(2S_{\lambda\lambda})\pi$ \\
		$\Xi_c'1D\left|\frac{3}{2}^+,2\right\rangle_{\lambda\lambda}$
		& 3178&
		& $\Xi_c\pi/\eta, \Xi'^{(*)}_c\pi, \Xi'_c\eta, \Sigma^{(*)}_c K, \Lambda_c\bar{K},\Lambda/\Sigma D,$ \\[-3pt]
		&&
		& $\Xi_c(1P_{\lambda/\rho})\pi, \Xi'_c(1P_\lambda/\rho)\pi,$ \\[-3pt]
		&&
		& $\Xi_c(2S_{\lambda\lambda})\pi, \Xi'_c(2S_{\lambda\lambda})\pi$ \\
		$\Xi_c'1D\left|\frac{5}{2}^+,2\right\rangle_{\lambda\lambda}$
		& 3185&
		& $\Xi_c\pi/\eta, \Xi'^{(*)}_c\pi, \Xi'_c\eta, \Sigma^{(*)}_c K, \Lambda_c\bar{K},\Lambda/\Sigma D,$ \\[-3pt]
		&&
		& $\Xi_c(1P_{\lambda/\rho})\pi, \Xi'_c(1P_\lambda/\rho)\pi,$ \\[-3pt]
		&&
		& $\Xi_c(2S_{\lambda\lambda})\pi, \Xi'_c(2S_{\lambda\lambda})\pi$ \\
		$\Xi_c'1D\left|\frac{5}{2}^+,3\right\rangle_{\lambda\lambda}$
		& 3169&$\Xi_c(3123)$
		& $\Xi_c\pi/\eta, \Xi'^{(*)}_c\pi, \Xi'_c\eta, \Sigma^{(*)}_c K, \Lambda_c\bar{K},\Lambda/\Sigma D,$ \\[-3pt]
		&&
		& $\Xi_c(1P_{\lambda/\rho})\pi, \Xi'_c(1P_\lambda/\rho)\pi,$ \\[-3pt]
		&&
		& $\Xi_c(2S_{\lambda\lambda})\pi, \Xi'_c(2S_{\lambda\lambda})\pi$ \\
		$\Xi_c'1D\left|\frac{7}{2}^+,3\right\rangle_{\lambda\lambda}$
		& 3158&
		& $\Xi_c\pi/\eta, \Xi'^{(*)}_c\pi, \Xi'_c\eta, \Sigma^{(*)}_c K, \Lambda_c\bar{K},\Lambda/\Sigma D,$ \\[-3pt]
		&&
		& $\Xi_c(1P_{\lambda/\rho})\pi, \Xi'_c(1P_\lambda/\rho)\pi,$ \\[-3pt]
		&&
		& $\Xi_c(2S_{\lambda\lambda})\pi, \Xi'_c(2S_{\lambda\lambda})\pi$ \\
		\hline
		$\Xi_c'1D\left|\frac{1}{2}^+,1\right\rangle_{\rho\rho}$
		& 3055&
		& $\Xi_c\pi/\eta, \Xi'^{(*)}_c\pi,\Sigma^{(*)}_c K, \Lambda_c\bar{K}$ \\[-3pt]
		&&
		& $\Xi_c(1P_{\lambda/\rho})\pi, \Xi'_c(1P_\lambda)\pi$ \\
		$\Xi_c'1D\left|\frac{3}{2}^+,1\right\rangle_{\rho\rho}$
		& 3236&
		& $\Xi_c\pi/\eta, \Xi'^{(*)}_c\pi/\eta,\Omega_c K,\Sigma^{(*)}_c K, \Lambda_c\bar{K}$ \\[-3pt]
		&&
		& $\Xi_c(1P_{\lambda/\rho})\pi, \Xi'_c(1P_\lambda/\rho)\pi, \Sigma_c(1P_\lambda)K,$ \\[-3pt]
		&&
		& $\Xi_c(2S_{\lambda\lambda/\rho\rho})\pi, \Xi'_c(2S_{\lambda\lambda/\rho\rho})\pi, \Xi_c(1D_{\lambda\lambda})\pi, \Xi_c'(1D_{\rho\rho})\pi$ \\
		$\Xi_c'1D\left|\frac{3}{2}^+,2\right\rangle_{\rho\rho}$
		& 3235&
		&  $\Xi_c\pi/\eta, \Xi'^{(*)}_c\pi/\eta,\Omega_c K,\Sigma^{(*)}_c K, \Lambda_c\bar{K}$ \\[-3pt]
		&&
		& $\Xi_c(1P_{\lambda/\rho})\pi, \Xi'_c(1P_\lambda/\rho)\pi, \Sigma_c(1P_\lambda)K,$ \\[-3pt]
		&&
		& $\Xi_c(2S_{\lambda\lambda/\rho\rho})\pi, \Xi'_c(2S_{\lambda\lambda/\rho\rho})\pi, \Xi_c(1D_{\lambda\lambda})\pi, \Xi_c'(1D_{\rho\rho})\pi$ \\
		$\Xi_c'1D\left|\frac{5}{2}^+,2\right\rangle_{\rho\rho}$
		& 3298&
		&  $\Xi_c\pi/\eta/\rho/\omega, \Xi'^{(*)}_c\pi/\eta,\Omega^{(*)}_c K,\Sigma^{(*)}_c K, \Lambda_c\bar{K}$ \\[-3pt]
		&&
		& $\Xi_c(1P_{\lambda/\rho})\pi, \Lambda_c(1P_{\rho})\bar{K}, \Xi'_c(1P_\lambda/\rho)\pi, \Sigma_c(1P_\lambda)K$ \\[-3pt]
		&&
		& $\Xi_c(2S_{\lambda\lambda/\rho\rho})\pi, \Lambda_c(2S_{\lambda\lambda})\bar{K}, \Xi'_c(2S_{\lambda\lambda/\rho\rho})\pi, \Xi_c(1D_{\lambda\lambda})\pi, \Xi_c'(1D_{\lambda\lambda/\rho\rho})\pi$ \\
		$\Xi_c'1D\left|\frac{5}{2}^+,3\right\rangle_{\rho\rho}$
		& 3263&
		& $\Xi_c\pi/\eta/\rho/\omega, \Xi'^{(*)}_c\pi/\eta,\Omega^{(*)}_c K,\Sigma^{(*)}_c K, \Lambda_c\bar{K}$\\[-3pt]
		&&
		& $\Xi_c(1P_{\lambda/\rho})\pi, \Xi'_c(1P_\lambda/\rho)\pi, \Sigma_c(1P_\lambda)K,$ \\[-3pt]
		&&
		& $\Xi_c(2S_{\lambda\lambda/\rho\rho})\pi, \Xi'_c(2S_{\lambda\lambda/\rho\rho})\pi, \Xi_c(1D_{\lambda\lambda})\pi, \Xi_c'(1D_{\rho\rho})\pi$ \\
		$\Xi_c'1D\left|\frac{7}{2}^+,3\right\rangle_{\rho\rho}$
		& 3300&
		& $\Xi_c\pi/\eta/\rho/\omega, \Xi'^{(*)}_c\pi/\eta,\Omega^{(*)}_c K,\Sigma^{(*)}_c K, \Lambda_c\bar{K}$ \\[-3pt]
		&&
		& $\Xi_c(1P_{\lambda/\rho})\pi, \Lambda_c(1P_{\rho})\bar{K}, \Xi'_c(1P_\lambda/\rho)\pi, \Sigma_c(1P_\lambda)K$ \\[-3pt]
		&&
		& $\Xi_c(2S_{\lambda\lambda/\rho\rho})\pi, \Lambda_c(2S_{\lambda\lambda})\bar{K}, \Xi'_c(2S_{\lambda\lambda/\rho\rho})\pi, \Xi_c(1D_{\lambda\lambda})\pi, \Xi_c'(1D_{\lambda\lambda/\rho\rho})\pi$ \\
		\hline
	\end{tabular}
	\label{tab:xicp_decay}
\end{table*}

\subsubsection{Strong decay of the $\Xi_c'$ resonances}
Within the framework of the QPC model, we investigate the strong decay properties of the $1P$- and $1D$-wave $\Xi_c'$ excited states. The corresponding quark model classification and possible two-body strong decay modes are listed in Table~\ref{tab:xicp_decay}.

\paragraph*{A. $1P$-wave states}

\begin{table*}[tp]
	\centering
	\caption{The strong decay properties of the $1P$-wave $\lambda$-mode $\Xi_c'$ states. The unit is MeV.}
	\label{tab:xicp_1Plambda_decay}
	\setlength{\tabcolsep}{2.8pt}
	\hspace{-10mm} 
	\fontsize{8.5}{8.5}\selectfont
	\begin{tabular}{l *{10}{c}}
		\hline\hline
		& \multicolumn{2}{c}{$\Xi_c'1P\big|J^P=\tfrac{1}{2}^-,0\big\rangle_\lambda$}
		& \multicolumn{2}{c}{$\Xi_c'1P\big|J^P=\tfrac{1}{2}^-,1\big\rangle_\lambda$}
		& \multicolumn{2}{c}{$\Xi_c'1P\big|J^P=\tfrac{3}{2}^-,1\big\rangle_\lambda$}
		& \multicolumn{2}{c}{$\Xi_c'1P\big|J^P=\tfrac{3}{2}^-,2\big\rangle_\lambda$}
		& \multicolumn{2}{c}{$\Xi_c'1P\big|J^P=\tfrac{5}{2}^-,2\big\rangle_\lambda$} \\[-4pt]
		& \multicolumn{2}{c}{\rule{2.0cm}{0.4pt}}
		& \multicolumn{2}{c}{\rule{2.0cm}{0.4pt}}
		& \multicolumn{2}{c}{\rule{2.0cm}{0.4pt}}
		& \multicolumn{2}{c}{\rule{2.0cm}{0.4pt}}
		& \multicolumn{2}{c}{\rule{2.0cm}{0.4pt}} \\[-2pt]
		Decay Width
		& \multicolumn{2}{c}{$2882$}
		& \multicolumn{2}{c}{$2867$}
		& $2930$ & $\Xi_c(2923)$
		& $2948$ & $\Xi_c(2939)$
		& $2969$ & $\Xi_c(2965)$ \\
		\hline
		$\Gamma[\Xi_c\pi]$
		&  \multicolumn{2}{c}{$30.91$} &  \multicolumn{2}{c}{$0.28$}  & $0.02$ &$0.02$ & $5.39$ & $5.12$ & $6.05$ & $6.00$ \\
		$\Gamma[\Xi_c'\pi]$
		&  \multicolumn{2}{c}{$0.33$}  & \multicolumn{2}{c}{ $23.79$} & $0.52$ & $0.46$& $1.18$ &$1.23$ & $0.66$ & $0.65$ \\
		$\Gamma[\Xi_c'^*\pi]$
		&  \multicolumn{2}{c}{$0.00$}   &  \multicolumn{2}{c}{$0.03$}  & $10.12$ & $9.93$ & $3.18$ & $3.03$& $0.67$ & $0.65$\\
		$\Gamma[\Sigma_c K]$
		& \multicolumn{2}{c}{ --}  &  \multicolumn{2}{c}{--} & -- & -- & $0.00$ & --& $0.02$ & $0.02$\\
		$\Gamma[\Lambda_c\bar{K}]$
		&  \multicolumn{2}{c}{$32.84$}  & \multicolumn{2}{c}{ $0.68$}  & $0.01$ & $0.01$ & $3.91$ & $3.63$ & $4.37$ & $4.32$\\
		$\Gamma\big[\Xi_c\big|J^P=\tfrac{1}{2}^-,1\big\rangle_\lambda \pi\big]$
		&  \multicolumn{2}{c}{--}  &  \multicolumn{2}{c}{--} & $0.00$ &-- & $0.13$ & $0.07$ & $0.00$ &-- \\
		$\Gamma\big[\Xi_c\big|J^P=\tfrac{3}{2}^-,1\big\rangle_\lambda \pi\big]$
		&  \multicolumn{2}{c}{--}  &  \multicolumn{2}{c}{--} & -- & --& -- & --& $0.09$ & $0.08$ \\
		\hline
		$\Gamma_{\text{Total}}$
		&  \multicolumn{2}{c}{$64.08$}   & \multicolumn{2}{c}{$24.78$}  & $10.67$ & $10.42$& $13.79$ & $13.08$& $11.86$ &$11.72$ \\
		\hline
		$\Gamma_{\text{Expt.}}$
		& \multicolumn{2}{c}{}
		& \multicolumn{2}{c}{}
		& \multicolumn{2}{c}{$5.3\pm0.9\pm1.4$}
		& \multicolumn{2}{c}{$14.8\pm8.8\pm2.5$}
		& \multicolumn{2}{c}{$14.1\pm0.9\pm1.3$} \\
		\hline\hline
	\end{tabular}
\end{table*}

For the $1P$-wave $\lambda$-mode excited $\Xi_c'$ baryons, there are five states: $\Xi_c'1P\big|J^P=1/2^-,0\big\rangle_\lambda$, $\Xi_c'1P\big|J^P=1/2^-,1\big\rangle_\lambda$, $\Xi_c'1P\big|J^P=3/2^-,1\big\rangle_\lambda$, $\Xi_c'1P\big|J^P=3/2^-,2\big\rangle_\lambda$, and $\Xi_c'1P\big|J^P=5/2^-,2\big\rangle_\lambda$. Fixing the masses of the five $1P$-wave $\lambda$-mode $\Xi_c'$ excited states to the values obtained from the previous mass-spectrum calculation, we calculated their decay properties, which are listed in Table~\ref{tab:xicp_1Plambda_decay}. From the table, $\Xi_c'1P\big|J^P=1/2^-,0\big\rangle_\lambda$ is likely to be a moderate state with a total decay width about $\Gamma\sim64$ MeV, and dominantly decays into the final states $\Xi_c\pi$ and $\Lambda_c\bar{K}$. The corresponding branching ratios are $48\%$ and $51\%$, respectively.
The total decay width of  $\Xi_c'1P\big|J^P=1/2^-,1\big\rangle_\lambda$ is slightly narrower, around $\Gamma\sim25$ MeV, and is almost entirely saturated by the decay channel $\Xi_c'\pi$.

We note that in our mass-spectrum analysis, the masses of $\Xi_c'1P\big|J^P=1/2^-,0\big\rangle_\lambda$ and $\Xi_c'1P\big|J^P=1/2^-,1\big\rangle_\lambda$ are both close to that of the experimentally observed $\Xi_c(2882)$.
Based on the strong decay properties, the $\Lambda_c\bar{K}$ is the dominate decay mode of the state $\Xi_c'\big|J^P=\tfrac12^-,0\big\rangle_\lambda$, with a large branching fraction, which is qualitatively consistent with the experimental observation that $\Xi_c(2882)$ is reconstructed in the $\Lambda_c\bar{K}$ channel. Nevertheless, our calculated total decay width is substantially larger than the experimental value. Therefore, we disfavor assigning $\Xi_c(2882)$ to this state. In contrast, the total width of $\Xi_c'\big|J^P=1/2^-,1\big\rangle_\lambda$ is roughly twice the experimental total decay width, which seems to be in rough agreement. While $\Xi_c(2882)$ is observed in the $\Lambda_c\bar{K}$ channel, the partial width of the $\Lambda_c\bar{K}$ mode for this state is found to be very small in our calculation. Accordingly, we also rule out this state as a candidate for $\Xi_c(2882)$.

For the state $\Xi_c'1P\big|J^P=3/2^-,1\big\rangle_\lambda$, the total decay width is about $\Gamma\sim11$ MeV and the dominant decay mode is $\Xi_c'^*\pi$. We examine its possibility as the $\Xi_c(2923)$ resonance in our mass-spectrum analysis, and adopt the physical mass $2923$ MeV to evaluate its strong decay properties. The resulting total decay width reads
\[
\Gamma_{\text{Total}} \simeq 10.42\,\text{MeV},
\]
which is about twice the experimental total width. Nevertheless, we argue that such a discrepancy is moderate and acceptable. Meanwhile, our decay calculation shows that the $\Xi_c'^{*}\pi$ mode dominates the decays, with its branching fraction given by
\[
\frac{\Gamma\left[\Xi_c'1P\big|J^P=3/2^-,1\big\rangle_\lambda \to \Xi_c'^{*}\pi\right]}{\Gamma_{\text{Total}}} \simeq 95\%.
\]
The subsequent decay $\Xi_c'^{*}\to \Xi_c\pi$ then yields the final $\Xi_c\pi\pi$ signature, which agrees with the experimental observation. However, while $\Xi_c(2923)$ has been observed in decay channel $\Lambda_c\bar{K}$ as well, we find that the corresponding decay channel is quite small in our model. This fact suggests that additional theoretical ingredients may be required for a full understanding. In summary, we tentatively identify $\Xi_c(2923)$ with the $\Xi_c'1P\big|J^P=3/2^-,1\big\rangle_\lambda$ state.

As for the state $\Xi_c'1P\big|J^P=3/2^-,2\big\rangle_\lambda$,  the decay width is approximately 14 MeV, with the dominant decay modes being $\Xi_c\pi$, $\Xi_c'^*\pi$ and $\Lambda_c\bar{K}$. We regard this state as a promising candidate for $\Xi_c(2939)$ from the mass perspective. Hence, we further fix the mass of  $\Xi_c'1P\big|J^P=3/2^-,2\big\rangle_\lambda$ on the physical mass 2942 MeV, and list its decay properties in Table~\ref{tab:xicp_1Plambda_decay} as well. The obtained total decay width is
\[
\Gamma_{\text{Total}} \simeq 13.08\,\text{MeV},
\]
which is in excellent agreement with the experimentally observed total width of state $\Xi_c(2939)$. Moreover,
the LHCb Collaboration reported the observation of $\Xi_c(2939)$ using the $\Lambda_c\bar{K}$ sample. In our decay analysis, the $\Lambda_c\bar{K}$ channel carries a considerable partial width, and the corresponding branching ratio is
\[
\frac{\Gamma\big[\Xi_c'1P\big|J^P=3/2^-,2\big\rangle_\lambda \to \Lambda_c\bar{K}\big]}{\Gamma_{\text{Total}}} \simeq 28\%.
\]
 This is also consistent with the decay properties of state  $\Xi_c(2939)$.
In addition, The decay channels $\Xi_c\pi$ and $\Xi_c'^{*}\pi$ also have substantial branching fractions, which are
\[
\frac{\Gamma\big[\Xi_c'1P\big|J^P=3/2^-,2\big\rangle_\lambda \to \Xi_c\pi\big]}{\Gamma_{\text{Total}}} \simeq 39\%
\]
and
\[
\frac{\Gamma\big[\Xi_c'1P\big|J^P=3/2^-,2\big\rangle_\lambda \to \Xi_c'^{*}\pi\big]}{\Gamma_{\text{Total}}} \simeq 23\%,
\]
respectively. Such considerable branching ratios suggest that, if state $\Xi_c(2939)$ can indeed be explained as the quark configuration $\Xi_c'1P\big|J^P=3/2^-,2\big\rangle_\lambda$, then state $\Xi_c(2939)$ is very likely to be observed experimentally in the final states $\Xi_c\pi$ and $\Xi_c'^{*}\pi$.

For the state $\Xi_c'1P\big|J^P=5/2^-,2\big\rangle_\lambda$, the total decay width reads $\Gamma\sim12$ MeV. The dominant decay channels are $\Xi_c\pi$ and $\Lambda_c\bar{K}$.
Since its predicted mass is $2969$ MeV, it is a promising candidate for $\Xi_c(2965)$ in our mass-spectrum analysis. We therefore adopt the physical mass $2965$ MeV to evaluate its decay width. $\Xi_c(2965)$ was discovered by the LHCb Collaboration in 2020 via the $\Lambda_c\bar{K}$ channel~\cite{LHCb:2020iby}, with the measured total width $\Gamma\simeq 14.1$ MeV. Our theoretical predicted total decay width is
\[
\Gamma_{\text{Total}} \simeq 11.72\,\text{MeV},
\]
which is in good agreement with the experimental value.
Moreover, the $\Lambda_c\bar{K}$ decay mode carries a substantial branching fraction, which is
\[
\frac{\Gamma\big[\Xi_c\big|J^P=\tfrac52^-,2\big\rangle_\lambda \to \Lambda_c\bar{K}\big]}{\Gamma_{\text{Total}}} \simeq 36\%.
\]
Such a sizable branching ratio is also consistent with the experimental fact that state $\Xi_c(2965)$ has been observed in decay channel $\Lambda_c\bar{K}$.

Additionally, the $\Xi_c\pi$ decay mode also dominates the decay pattern as listed in the table, with the corresponding branching fraction being
\[
\frac{\Gamma\big[\Xi_c'1P\big|J^P=5/2^-,2\big\rangle_\lambda \to \Xi_c\pi\big]}{\Gamma_{\text{Total}}} \simeq 50\%.
\]
This suggests that the $\Xi_c\pi$ decay channel is a promising additional mode for future experimental searches for $\Xi_c(2965)$.

\begin{table}[htb]
	\centering
	\caption{The strong decay properties of the $1P$-wave $\rho$-mode $\Xi_c'$ states.The unit is MeV.}
	\label{tab:xicp_1Prho_decay}
	\setlength{\tabcolsep}{2.8pt}
	\fontsize{8.5}{8.5}\selectfont
	\begin{tabular}{l *{4}{c}}
		\hline\hline
		& $\Xi_c'1P\big|J^P=\tfrac{1}{2}^-,1\big\rangle_\rho$
		& $\Xi_c'1P\big|J^P=\tfrac{3}{2}^-,1\big\rangle_\rho$ \\[-4pt]
		& \rule{2.0cm}{0.4pt}
		&\rule{2.0cm}{0.4pt} \\[-2pt]
		Decay width & $2946$  & $2976$ \\
		\hline
		$\Gamma[\Xi_c'\pi]$
		& $15.54$  & $2.13$  \\
		$\Gamma[\Xi_c'^*\pi]$
		& $0.92$ & $19.61$  \\
		$\Gamma[\Sigma_c K]$
		& -- & $0.09$  \\
		$\Gamma\big[\Xi_c\big|J^P=\tfrac{1}{2}^-,0\big\rangle_\rho \pi\big]$
		& --& $0.00$ \\
		\hline
		$\Gamma_{\text{Total}}$
		& $16.46$  & $21.83$ \\
	\hline\hline
	\end{tabular}
\end{table}

For the two $1P$-wave $\rho$-mode $\Xi_c'$ baryons $\Xi_c'1P\big|J^P=1/2^-,1\big\rangle_\rho$ and $\Xi_c'1P\big|J^P=3/2^-,1\big\rangle_\rho$, their decay properties are collected in Table~\ref{tab:xicp_1Prho_decay}. The state $\Xi_c'1P\big|J^P=1/2^-,1\big\rangle_\rho$ is likely to be a relatively narrow state with a width of approximately $\Gamma\sim16$ MeV. The dominant decay channel is $\Xi_c'\pi$, with branching ratio
\[
\frac{\Gamma\big[\Xi_c'\big|J^P=\tfrac12^-,1\big\rangle_\rho \to \Xi_c'\pi\big]}{\Gamma_{\text{Total}}} \simeq 94\%.
\]
These decay properties suggest that this state is highly likely to be observed in decay channel $\Xi_c'\pi$ in future experiments.

For $\Xi_c'\big|J^P=\tfrac32^-,1\big\rangle_\rho$,  the total decay width is approximately 22 MeV, and the dominant decay mode is $\Xi_c'^*\pi$, with a branching ratio of
\[
\frac{\Gamma\big[\Xi_c'1P\big|J^P=3/2^-,1\big\rangle_\rho \to \Xi_c'^*\pi\big]}{\Gamma_{\text{Total}}} \simeq 90\%.
\]
Such a sizable branching ratio suggests that this state is highly likely to be observed in the final state $\Xi_c\pi\pi$ via the decay chain $\Xi_c'1P\big|J^P=3/2^-,1\big\rangle_\rho\rightarrow \Xi_c'^*\pi\rightarrow \Xi_c\pi\pi$.

We notice that based on our mass-spectrum analysis, $\Xi_c'1P\big|J^P=1/2^-,1\big\rangle_\rho$ and $\Xi_c'1P\big|J^P=3/2^-,1\big\rangle_\rho$ were initially considered as candidates for $\Xi_c(2939)$ and $\Xi_c(2965)$, respectively. However, according to our decay properties, although the theoretical total widths are of the same order as those of the experimental states,  $\Xi_c(2939)$ and $\Xi_c(2965)$ were observed in decay channel $\Lambda_c\bar{K}$, whereas our calculation shows that this channel is forbidden, i.e., very small. Therefore, we do not consider $\Xi_c'1P\big|J^P=1/2^-,1\big\rangle_\rho$ and $\Xi_c'1P\big|J^P=3/2^-,1\big\rangle_\rho$ as the candidates for $\Xi_c(2939)$ and $\Xi_c(2965)$, respectively, for the time being.

\paragraph*{B. $1D$-wave states}

\vspace{\baselineskip}\vspace{\baselineskip}
\begin{table*}[tp]
	\centering
	\small
	\caption{The strong decay properties of the $1D$-wave $\lambda$-mode $\Xi_c'$ states.The unit is MeV.}
	\label{tab:xicp_1Dlambdalambda_decay}
	\setlength{\tabcolsep}{1.5pt}
	\hspace{-10 mm}  
	\fontsize{8.5}{9}\selectfont
	\renewcommand{\arraystretch}{1.4}
	\begin{tabular}{l *{12}{c}}
		\hline\hline
		& \multicolumn{2}{c}{$\Xi_c'1D\big|J^P=\tfrac{1}{2}^+,1\big\rangle_{\lambda\lambda}$}
		& \multicolumn{1}{c}{$\Xi_c'1D\big|J^P=\tfrac{3}{2}^+,1\big\rangle_{\lambda\lambda}$}
		& \multicolumn{1}{c}{$\Xi_c'1D\big|J^P=\tfrac{3}{2}^+,2\big\rangle_{\lambda\lambda}$}
		& \multicolumn{1}{c}{$\Xi_c'1D\big|J^P=\tfrac{5}{2}^+,2\big\rangle_{\lambda\lambda}$}
		& \multicolumn{2}{c}{$\Xi_c'1D\big|J^P=\tfrac{5}{2}^+,3\big\rangle_{\lambda\lambda}$}
		& \multicolumn{1}{c}{$\Xi_c'1D\big|J^P=\tfrac{7}{2}^+,3\big\rangle_{\lambda\lambda}$}\\[-4pt]
		& \multicolumn{2}{c}{\rule{2.5cm}{0.4pt}}
		& \multicolumn{1}{c}{\rule{2.0cm}{0.4pt}}
		& \multicolumn{1}{c}{\rule{2.0cm}{0.4pt}}
		& \multicolumn{1}{c}{\rule{2.0cm}{0.4pt}}
		& \multicolumn{2}{c}{\rule{2.5cm}{0.4pt}}
		& \multicolumn{1}{c}{\rule{2.0cm}{0.4pt}} \\[-2pt]
		Decay Width
		& $3153$ & $\Xi_c(3123)$
		& $3181$
		& $3178$
		& $3185$
		& $3169$ & $\Xi_c(3123)$
		& $3158$ \\
		\hline
		$\Gamma[\Xi_c\pi/\eta]$
		& $0.15/0.02$ &$0.14/0.01$
		& $3.25/0.96$
		& $0.00/0.03$
		& $0.30/0.17$
		& $2.02/0.17$  &$1.65/0.09$
		& $0.09/0.00$ \\
		$\Gamma[\Xi_c'\pi/\eta]$
		&$0.04/0.00$ &$0.03/-$
		& $0.39/0.07$
		& $2.59/0.23$
		& $0.69/0.02$
		& $0.09/0.01$ &$0.04/-$
		& $0.02/0.00$ \\
		$\Gamma[\Xi_c'^*\pi]$
		&$0.10$ &$0.09$
		& $0.31$
		& $1.64$
		& $2.42$
		& $0.88$  &$0.83$
		& $0.06$ \\
		$\Gamma[\Sigma_c K]$
		& $0.06$ &$0.05$
		& $1.09$
		& $5.94$
		& $1.31$
		& $0.06$ &$0.01$
		& $0.04$ \\
		$\Gamma[\Sigma_c^* K]$
		& $0.16$ &$0.12$
		& $0.81$
		& $3.20$
		& $5.85$
		& $2.11$ &$1.66$
		& $0.04$ \\
		$\Gamma[\Lambda_c\bar{K}]$
		& $0.11$ &$0.10$
		& $2.27$
		& $0.01$
		& $0.21$
		& $1.75$  &$1.43$
		& $0.07$ \\
		$\Gamma[\Lambda/\Sigma D]$
		&$3.13/2.07$ &$2.70/1.36$
		& $1.09/0.93$
		& $10.34/8.77$
		& $0.50/0.17$
		& $0.00/0.00$ &$0.00/0.00$
		& $0.14/0.03$ \\
		$\Gamma\big[\Xi_c 1P\big|J^P=\tfrac12^-,0/1\big\rangle_\rho \pi\big]$
		&$0.00/0.12$ &$0.00/0.10$
		& $0.02/0.00$
		& $0.00/0.02$
		& $0.11/0.00$
		& $0.01/0.09$ &$0.01/0.05$
		& $0.00/0.00$ \\
		$\Gamma\big[\Xi_c 1P\big|J^P=\tfrac32^-,1/2\big\rangle_\rho \pi\big]$
		& $0.00/0.00$ &$-/-$
		& $0.76/0.01$
		& $0.44/1.16$
		& $0.00/0.00$
		& $0.00/0.00$ &$-/-$
		& $0.00/0.00$ \\
		$\Gamma\big[\Xi_c 1P\big|J^P=\tfrac12^-/\tfrac32^-,1\big\rangle_\lambda \pi\big]$
		& $2.57/1.57$ &$2.55/1.44$
		& $0.08/29.75$
		& $0.35/0.20$
		& $0.45/0.16$
		& $1.16/0.03$ &$0.58/0.02$
		& $0.17/0.00$ \\
		$\Gamma\big[\Xi_c' 1P\big|J^P=\tfrac12^-/\tfrac32^-,1\big\rangle_\rho \pi\big]$
		& $0.05/0.01$ &$0.03/0.00$
		& $0.00/0.39$
		& $0.00/0.00$
		& $0.00/0.00$
		&$0.02/0.02$ &$0.01/0.01$
		& $0.00/0.01$ \\
		$\Gamma\big[\Xi_c' 1P\big|J^P=\tfrac12^-,0/1\big\rangle_\lambda \pi\big]$
		&$0.00/2.58$ &0.00/2.23
		& $0.16/0.04$
		& $0.08/0.13$
		& $0.02/0.05$
		& $0.04/0.11$ &$0.01/0.04$
		& $0.00/0.00$ \\
		$\Gamma\big[\Xi_c' 1P\big|J^P=\tfrac32^-,1/2\big\rangle_\lambda \pi\big]$
		&$0.00/0.00$ &0.00/0.00
		& $12.86/0.07$
		& $7.39/20.35$
		& $0.04/0.00$
		& $0.00/0.05$ &$0.00/0.00$
		& $0.00/0.00$ \\
		$\Gamma\big[\Xi_c' 1P\big|J^P=\tfrac52^-,2\big\rangle_\lambda \pi\big]$
		&$0.00$ &$0.00$
		& $0.01$
		& $0.00$
		& $23.44$
		& $0.09$ &$0.04$
		& $0.00$ \\
		$\Gamma\big[\Xi_c 2S\big|J^P=\tfrac12^+,1\big\rangle_{\lambda\lambda} \pi \big]$
		&$0.05$ &$0.01$
		& $18.24$
		& $0.00$
		& $0.00$
		& $0.00$ &$0.00$
		& $0.00$ \\
		$\Gamma\big[\Xi_c' 2S\big|J^P=\tfrac12^+/\tfrac32^+,1\big\rangle_{\lambda\lambda} \pi \big]$
		&$0.00$/-- &$0.00/-$
		& $0.00/0.00$
		& $0.03/0.00$
		& $0.00/0.00$
		& $0.00/0.00$ &$0.00/-$
		& $0.00$/-- \\
		\hline
		$\Gamma_{\text{Total}}$
		& $12.79$&$10.96$
		&$73.56$
		&$62.90$
		& $35.91$
		& $8.71$ &$6.48$
		& $0.67$ \\
		\hline
		$\Gamma_{\text{Expt.}}$
		& \multicolumn{2}{c}{$4.4\pm3.4\pm1.7$}
		&
		&
		&
		& \multicolumn{2}{c}{$4.4\pm3.4\pm1.7$}
		& \\
		\hline\hline
	\end{tabular}
\end{table*}

For the six $\lambda$-mode excited $\Xi_c'$ states, their decay properties are listed in Table~\ref{tab:xicp_1Dlambdalambda_decay}. It is shown that the state $\Xi_c'1D\big|J^P=1/2^+,1\big\rangle_{\lambda\lambda}$ has a total width of $\Gamma\sim13$ MeV. Decay channels $\Lambda D$ and $\Sigma D$, decaying into single heavy-flavor meson final states, have sizable branching ratios of 24$\%$ and 16$\%$, respectively. Meanwhile, the decay channels with a P-wave state and a single pion in the final state also carry non-negligible branching ratios; for instance, the branching ratios of the decay channels $\Xi_c 1P\big|J^P=1/2^-,1\big\rangle_\lambda \pi$ and $\Xi_c' 1P\big|J^P=1/2^-,1\big\rangle_\lambda \pi$ both amount to 20$\%$.

In our mass-spectrum analysis,  the mass of state $\Xi_c(3123)$ lies in the mass region of the $1D$-wave $\lambda$-mode state of $\Xi_c'$. Hence, We further adopt the physical mass of $\Xi_c(3123)$ to evaluate its decay widths. The predicted total decay width is
\[
\Gamma_{\text{Total}} \simeq 10.96\,\text{MeV},
\]
which is close to the upper limit on the total width of the observed state $\Xi_c(3123)$. However, we note that the branching ratio of decay channel $\Sigma_c^*K$ is quite small, about
\[
\frac{\Gamma\big[\Xi_c'1D\big|J^P=1/2^+,1\big\rangle_{\lambda\lambda} \to \Sigma_c^* K\big]}{\Gamma_{\text{Total}}} \simeq 1\%,
\]
 which is inconsistent with the observation of state $\Xi_c(3123)$ by the BABAR Collaboration in 2008 from the mass spectrum of $\Sigma_c(2520)^{++}K^- \to \Lambda^+ K^- \pi^+$~\cite{BaBar:2007zjt}. In 2014, however, the Belle Collaboration did not reconstruct state $\Xi_c(3123)$ in this channel~\cite{Belle:2013htj}. Therefore, in view of the limited experimental data on state $\Xi_c(3123)$, we cannot yet exclude the possibility that state $\Xi_c(3123)$ is a candidate for the quark configuration $\Xi_c'1D\big|J^P=1/2^+,1\big\rangle_{\lambda\lambda}$.

The two spin-parity $J^P=3/2^-$ states $\Xi_c'1D\big|J^P=3/2^+,1\big\rangle_{\lambda\lambda}$ and $\Xi_c'1D\big|J^P=3/2^+,2\big\rangle_{\lambda\lambda}$ have comparable decay widths of about (60–70) MeV. However, their dominant decay modes are quite different. For state $\Xi_c'1D\big|J^P=3/2^+,1\big\rangle_{\lambda\lambda}$, the main decay channel is $\Xi_c1P\big|J^P=3/2^-,1\big\rangle_{\lambda}\pi$, with a branching fraction of 40$\%$; channels $\Xi_c2S\big|J^P=1/2^+,1\big\rangle_{\lambda\lambda}\pi$ and $\Xi_c'1P\big|J^P=3/2^-,1\big\rangle_{\lambda}\pi$  also have non-negligible branching ratios of 25$\%$ and 17$\%$, respectively. For state $\Xi_c'1D\big|J^P=3/2^+,2\big\rangle_{\lambda\lambda}$, the main decay channel is
$\Xi_c'1P\big|J^P=3/2^-,2\big\rangle_{\lambda}\pi$, with a branching ratio of 32$\%$; the branching ratios of channels $\Lambda D$, $\Sigma D$ and $\Xi_c'1P\big|J^P=3/2^-,1\big\rangle_{\lambda}\pi$ are also sizable, at 16$\%$, 14$\%$ and 12$\%$, respectively. In summary, the decay channels with a $P$-wave state plus a single pion in the final state play a dominant role among the decay modes of these two states and deserve attention in future experiments.

The state $\Xi_c'1D\big|J^P=5/2^+,2\big\rangle_{\lambda\lambda}$ may be a moderate state with a decay width of about $\Gamma\sim36$ MeV. It mainly decays into the channel $\Xi_c'1P\big|J^P=5/2^-,2\big\rangle_{\lambda}\pi$, and the corresponding branching ratio can reach up to 65$\%$. Consequently, this state has a good chance of being observed in the final states $\Xi_c\pi\pi$ and $\Lambda_c \bar{K}\pi$ via the decay chains $\Xi_c'1D\big|J^P=5/2^+,2\big\rangle_{\lambda\lambda}\rightarrow \Xi_c'1P\big|J^P=5/2^-,2\big\rangle_{\lambda}\pi\rightarrow\Xi_c\pi(\Lambda_c \bar{K})\pi$.

For state $\Xi_c'1D\big|J^P=5/2^+,3\big\rangle_{\lambda\lambda}$, it is very likely to be a narrow state, with a width of only a few MeV. The dominant decay channels are $\Sigma_c^*K$, $\Xi_c\pi$ and $\Lambda_c\bar{K}$. Given the decay properties of the observed state $\Xi_c(3123)$, state $\Xi_c'1D\big|J^P=5/2^+,3\big\rangle_{\lambda\lambda}$ is highly likely to be a candidate for $\Xi_c(3123)$. We therefore fix the mass of state $\Xi_c'1D\big|J^P=5/2^+,3\big\rangle_{\lambda\lambda}$ at its physical mass of 3123 MeV and calculate its decay properties, which are listed in Table~\ref{tab:xicp_1Dlambdalambda_decay}.

As can be seen from the table, the total width of $\Xi_c'1D\big|J^P=5/2^+,3\big\rangle_{\lambda\lambda}$ is
\[
\Gamma_{\text{Total}} \simeq 6.48\,\text{MeV}.
\]
Within the uncertainties, this value is in excellent agreement with the total decay width of $\Xi_c(3123)$. The dominant decay channels $\Xi_c\pi$ and $\Sigma_c^*K$ have branching ratios of
\[
\frac{\Gamma\big[\Xi_c'1D\big|J^P=5/2^+,3\big\rangle_{\lambda\lambda} \to \Xi_c\pi\big]}{\Gamma_{\text{Total}}} \simeq 25\%
\]
and
\[
\frac{\Gamma\big[\Xi_c'1D\big|J^P=5/2^+,3\big\rangle_{\lambda\lambda} \to \Sigma_c^* K\big]}{\Gamma_{\text{Total}}} \simeq 26\%,
\]
respectively. In addition, the branching ratio of decay channel $\Lambda_c\bar{K}$ is also considerable, amounting to
\[
\frac{\Gamma\big[\Xi_c'1D\big|J^P=5/2^+,3\big\rangle_{\lambda\lambda} \to \Lambda_c\bar{K}\big]}{\Gamma_{\text{Total}}} \simeq 22\%.
\]
Therefore,  if $\Xi_c(3123)$ is indeed the candidate for state $\Xi_c'1D\big|J^P=5/2^+,3\big\rangle_{\lambda\lambda}$, then apart from the already observed decay channel $\Sigma_c^*K\rightarrow \Lambda_c\pi K$, experimental collaborations could also search for $\Xi_c(3123)$ signals in decay channels $\Xi_c\pi$ and $\Lambda_c\bar{K}$.

The state $\Xi_c'1D\big|J^P=7/2^+,3\big\rangle_{\lambda\lambda}$ is very likely to be a very narrow state, with a total width of only a few tenths of an MeV. It decays predominantly into the $\Xi_c1P\big|J^P=1/2^-,1\big\rangle_{\lambda}\pi$ and $\Lambda D$ final states, with branching ratios of 25$\%$ and 21$\%$, respectively. Therefore, it can be observed experimentally in the $\Lambda D$ decay channel or, via cascade decays $\Xi_c'1D\big|J^P=7/2^+,3\big\rangle_{\lambda\lambda}\rightarrow \Xi_c1P\big|J^P=1/2^-,1\big\rangle_{\lambda}\pi\rightarrow \Xi_c'\pi\pi$, in the $\Xi_c'\pi\pi$ decay channel.

In summary, the $1D$-wave $\lambda$-mode $\Xi_c'$ baryons are relatively narrow and are likely to be observed in future experiments. Moreover, $\Xi_c'1D\big|J^P=1/2^+,1\big\rangle_{\lambda\lambda}$ and $\Xi_c'1D\big|J^P=5/2^+,3\big\rangle_{\lambda\lambda}$ are candidates for the observed state $\Xi_c(3123)$.  By comparing the decay properties with those of $\Xi_c(3123)$, candidate $\Xi_c'1D\big|J^P=5/2^+,3\big\rangle_{\lambda\lambda}$ is found to be more favorable.

\begin{table*}[tp]
	\centering
	\caption{The strong decay properties of the $1D$-wave $\rho$-mode $\Xi_c'$ states. The unit is MeV.}
	\label{tab:xicp_1Drhorho_decay}
	\setlength{\tabcolsep}{1.0pt}
	\hspace{-10mm}  
	\fontsize{8.5}{8.5}\selectfont
	\renewcommand{\arraystretch}{1.4}  
	\begin{tabular}{l *{12}{c}}
		\hline
		& \multicolumn{2}{c}{$\Xi_c'1D\big|J^P=\tfrac{1}{2}^+,1\big\rangle_{\rho\rho}$}
		& \multicolumn{2}{c}{$\Xi_c'1D\big|J^P=\tfrac{3}{2}^+,1\big\rangle_{\rho\rho}$}
		& \multicolumn{2}{c}{$\Xi_c'1D\big|J^P=\tfrac{3}{2}^+,2\big\rangle_{\rho\rho}$}
		& \multicolumn{2}{c}{$\Xi_c'1D\big|J^P=\tfrac{5}{2}^+,2\big\rangle_{\rho\rho}$}
		& \multicolumn{2}{c}{$\Xi_c'1D\big|J^P=\tfrac{5}{2}^+,3\big\rangle_{\rho\rho}$}
		& \multicolumn{2}{c}{$\Xi_c'1D\big|J^P=\tfrac{7}{2}^+,3\big\rangle_{\rho\rho}$}\\
		\cmidrule(lr){2-3} \cmidrule(lr){4-5} \cmidrule(lr){6-7}
		\cmidrule(lr){8-9} \cmidrule(lr){10-11} \cmidrule(lr){12-13}
		Decay Width& $M=3055$ && $M=3236$ && $M=3235$ && $M=3298$ && $M=3263$ && $M=3300$ \\
		\hline
		$\Gamma[\Xi_c\pi/\eta]$
		& $0.00/0.00$ & & $11.35/4.12$ & & $0.01/0.18$ & & $0.09/0.17$ & & $3.86/0.74$ & & $0.52/0.21$ & \\
		$\Gamma[\Xi_c'\pi/\eta]$
		& $0.00$/-- & & $1.97/0.68$ & & $10.23/2.10$ & & $1.55/0.24$ & & $0.68/0.00$ & & $0.20/0.06$ & \\
		$\Gamma[\Xi_c'^*\pi/\eta]$
		& $0.00$/-- & & $0.98/0.11$ & & $6.49/0.34$ & & $3.45/1.16$ & & $1.52/0.39$ & & $2.85/0.05$ & \\
		$\Gamma[\Xi_c\rho/\omega]$
		& --/-- & & --/-- & & --/-- & & $0.62/0.19$ & & $0.07/0.01$ & & $0.30/0.09$ & \\
		$\Gamma[\Omega_c/\Omega_c^* K]$
		& --/--  & & $0.35$/--  & & $0.92$/-- & & $0.13/0.45$ & & $0.00/0.01$ & & $0.04/0.00$ & \\
		$\Gamma[\Sigma_c/\Sigma_c^* K]$
		& $0.00/0.00$ & & $5.43/2.45$ & & $24.21/13.65$ & & $3.99/9.96$ & & $1.45/4.71$ & & $0.53/5.86$ & \\
		$\Gamma[\Lambda_c\bar{K}]$
		& $0.00$ & & $7.49$ & & $0.00$ & & $0.03$ & & $3.42$ & & $0.37$ & \\
		$\Gamma\big[\Xi_c\ 1P\big|J^P=\tfrac{1}{2}^-,0/1\big\rangle_\rho \pi\big]$
		& $0.00/0.00$ & & $2.00/0.19$ & & $0.05/2.17$ & & $0.17/0.53$ & & $0.52/0.06$ & & $0.00/0.72$ & \\
		$\Gamma\big[\Xi_c\ 1P\big|J^P=\tfrac{3}{2}^-,1/2\big\rangle_\rho \pi\big]$
		& --/--& & $9.59/9.77$ & & $4.89/4.99$ & & $2.00/0.49$ & & $0.45/0.14$ & & $0.21/0.09$ & \\
		$\Gamma\big[\Xi_c\ 1P\big|J^P=\tfrac{5}{2}^-,2\big\rangle_\rho \pi\big]$
		& --& & $0.30$ & & $0.07$ & & $11.10$ & & $0.00$ & & $0.05$ & \\
		$\Gamma\big[\Lambda_c\ 1P\big|J^P=\tfrac{1}{2}^-,0\big\rangle_\rho \bar{K}\big]$
		&-- & & -- & & -- & & $0.65$ & & -- & & $0.00$ & \\
		$\Gamma\big[\Xi_c\ 1P\big|J^P=\tfrac{1}{2}^-/\tfrac{3}{2}^-,1\big\rangle_\lambda \pi\big]$
		& $0.00/0.00$ & & $0.14/5.39$ & & $0.00/0.07$ & & $0.00/0.00$ & & $0.33/0.19$ & & $0.09/0.96$ & \\
		$\Gamma\big[\Xi_c'\ 1P\big|J^P=\tfrac{1}{2}^-/\tfrac{3}{2}^-,1\big\rangle_\rho \pi\big]$
		&--/-- & & $5.56/5.92$ & & $0.02/0.00$ & & $0.74/0.64$ & & $0.64/2.79$ & & $0.00/0.43$ & \\
		$\Gamma\big[\Xi_c'\ 1P\big|J^P=\tfrac{1}{2}^-,0/1\big\rangle_\lambda \pi\big]$
		& $0.00/0.00$ & & $0.01/0.00$ & & $0.00/0.02$ & & $0.23/0.02$ & & $0.00/0.12$ & & $0.00/0.01$ & \\	
		$\Gamma\big[\Xi_c'\ 1P\big|J^P=\tfrac{3}{2}^-,1/2\big\rangle_\lambda \pi\big]$
		& --/--& & $1.28/0.02$ & & $0.68/2.10$ & & $0.05/0.08$ & & $0.04/0.01$ & & $0.21/0.01$ & \\	
		$\Gamma\big[\Xi_c'\ 1P\big|J^P=\tfrac{5}{2}^-,2\big\rangle_\lambda \pi\big]$
		&--/-- & & $0.01$ & & $0.00$ & & $0.76$ & & $0.00$ & & $0.03$ & \\	
		$\Gamma\big[\Sigma_c\ 1P\big|J^P=\tfrac{1}{2}^-,0/1\big\rangle_\lambda \ K\big]$
		&--/-- & & --/$0.00$ & & --/$0.00$ & & $0.00/0.05$ & & --/$0.03$ & & $0.00/0.00$ & \\	
		$\Gamma\big[\Sigma_c\ 1P\big|J^P=\tfrac{3}{2}^-,1/2\big\rangle_\lambda \ K\big]$
		&--/-- & &  --/-- & &  --/-- & & $0.00/0.02$ & & --/-- & & $0.00/0.00$ & \\	
		$\Gamma\big[\Sigma_c\ 1P\big|J^P=\tfrac{5}{2}^-,2\big\rangle_\lambda \ K \big]$
		&-- & & -- & & -- & & $0.99$ & & -- & & $0.00$ & \\	
		$\Gamma\big[\Xi_c\ 2S\big|J^P=\tfrac{1}{2}^+,1\big\rangle_{\rho\rho} \pi \big]$
		& -- & & $0.02$ & & $0.00$ & & $0.00$ & & $0.00$ & & $0.00$ & \\	
		$\Gamma\big[\Xi_c\ 2S\big|J^P=\tfrac{1}{2}^+,1\big\rangle_{\lambda\lambda} \pi \big]$
		& -- & & $49.77$ & & $0.01$ & & $0.00$ & & $0.00$ & & $0.00$ & \\	
		$\Gamma\big[\Lambda_c\ 2S\big|J^P=\tfrac{1}{2}^+,1\big\rangle_{\lambda\lambda} \bar{K} \big]$
		& -- & & -- & & -- & & $0.00$ & & -- & & $0.00$ & \\	
		$\Gamma\big[\Xi_c'\ 2S\big|J^P=\tfrac{1}{2}^+/\tfrac{3}{2}^+,1\big\rangle_{\rho\rho} \pi \big]$
		& --/--& & $0.00$/-- & & $0.02$/-- & & $0.00/0.11$ & & $0.00/0.00$ & & $0.00/0.00$ & \\	
		$\Gamma\big[\Xi_c'\ 2S\big|J^P=\tfrac{1}{2}^+/\tfrac{3}{2}^+,1\big\rangle_{\lambda\lambda} \pi \big]$
		& --/--& & $0.00/0.04$ & & $0.10/0.00$ & & $0.00/0.32$ & & $0.00/0.00$ & & $0.00/0.00$ & \\	
		$\Gamma\big[\Xi_c\ 1D\big|J^P=\tfrac{3}{2}^+/\tfrac{5}{2}^+,2\big\rangle_{\lambda\lambda} \pi \big]$
		&--/-- & & $0.00/0.00$ & & $0.00/0.00$ & & $0.00/0.00$ & & $0.00/0.00$ & & $0.00/0.01$ & \\	
		$\Gamma\big[\Xi_c'\ 1D\big|J^P=\tfrac{1}{2}^+,1\big\rangle_{\rho\rho} \pi \big]$
		&-- & & $0.03$ & & $0.00$ & & $0.00$ & & $0.00$ & & $0.00$ & \\	
		$\Gamma\big[\Xi_c'\ 1D\big|J^P=\tfrac{1}{2}^+,1\big\rangle_{\lambda\lambda} \pi \big]$
		& --& & -- & & -- & & $0.00$ & & -- & & $0.00$ & \\	
		$\Gamma\big[\Xi_c'\ 1D\big|J^P=\tfrac{7}{2}^+,3\big\rangle_{\lambda\lambda} \pi \big]$
		&-- & & -- & & -- & & $0.00$ & & -- & & $0.00$ & \\	
		\hline
		$\Gamma_{\text{Total}}$
		& $0.00$& &$124.97$ & &$73.32$ & & $40.98$ & & $22.18$ & & $13.90$ & \\
		\hline
	\end{tabular}
\end{table*}

Similarly, we fix the masses of the six $1D$-wave $\rho$-mode excited $\Xi_c'$ states to the values obtained from our mass-spectrum calculation and give their strong decay properties, as listed in Table~\ref{tab:xicp_1Drhorho_decay}.

It is shown that $\Xi_c'1D\big|J^P=1/2^+,1\big\rangle_{\rho\rho}$ is probably a very narrow state, with a width below 1 MeV, while $\Xi_c'1D\big|J^P=3/2^+,1\big\rangle_{\rho\rho}$ is a relatively broad state, with a width exceeding 100 MeV. The dominant decay channel of $\Xi_c'1D\big|J^P=3/2^+,1\big\rangle_{\rho\rho}$ is $\Xi_c2S\big|J^P=1/2^+,1\big\rangle_{\lambda\lambda}\pi$, and the branching ratio can reach up to 40$\%$. Owing to its relatively broad width, observing the signal of the state $\Xi_c'1D\big|J^P=3/2^+,1\big\rangle_{\rho\rho}$ experimentally is rather challenging.

The total decay width of $\Xi_c'1D\big|J^P=3/2^+,2\big\rangle_{\rho\rho}$ is about $\Gamma\sim73$ MeV, and the mainly decay channels are $\Sigma_c^{(*)}K$ and $\Xi_c'\pi$. The predicted branching ratios are
\[
\frac{\Gamma\big[\Xi_c'1D\big|J^P=3/2^+,2\big\rangle_{\rho\rho} \to \Sigma_cK\big]}{\Gamma_{\text{Total}}} \simeq 33\%,
\]
\[
\frac{\Gamma\big[\Xi_c'1D\big|J^P=3/2^+,2\big\rangle_{\rho\rho} \to \Sigma_c^*K\big]}{\Gamma_{\text{Total}}} \simeq 19\%
\]
and
\[
\frac{\Gamma\big[\Xi_c'1D\big|J^P=3/2^+,2\big\rangle_{\rho\rho} \to \Xi_c'\pi\big]}{\Gamma_{\text{Total}}} \simeq 14\%,
\]
respectively. This state may be observed in its main decay channels.

$\Xi_c'1D\big|J^P=5/2^+,2\big\rangle_{\rho\rho}$ is very likely to be a moderately broad state, with a width of about 40 MeV. It decays predominantly via the $\Sigma_c^*K$ channel, with a branching ratio of
\[
\frac{\Gamma\big[\Xi_c'1D\big|J^P=5/2^+,2\big\rangle_{\rho\rho} \to \Sigma_c^*K\big]}{\Gamma_{\text{Total}}} \simeq 24\%.
\]
Meanwhile, the decay channels $\Xi_c'^*\pi$ and $\Sigma_cK$ also have sizable branching ratios, which are
\[
\frac{\Gamma\big[\Xi_c'1D\big|J^P=5/2^+,2\big\rangle_{\rho\rho} \to \Xi_c'^*\pi\big]}{\Gamma_{\text{Total}}} \simeq 8\%
\]
and
\[
\frac{\Gamma\big[\Xi_c'1D\big|J^P=5/2^+,2\big\rangle_{\rho\rho} \to \Sigma_cK\big]}{\Gamma_{\text{Total}}} \simeq 10\%,
\]
respectively. Therefore, based on our theoretical results, state $\Xi_c'1D\big|J^P=5/2^+,2\big\rangle_{\rho\rho}$ is very likely to be observed experimentally in the final states $\Lambda_c\pi K$ and $\Xi_c\pi \pi$ via the decay chains $\Xi_c'1D\big|J^P=5/2^+,2\big\rangle_{\rho\rho}\rightarrow\Sigma_c^{(*)}K\rightarrow\Lambda_c\pi K$ and $\Xi_c'1D\big|J^P=5/2^+,2\big\rangle_{\rho\rho}\rightarrow\Xi_c'^*\pi\rightarrow\Xi_c\pi \pi$.

For $\Xi_c'1D\big|J^P=5/2^+,3\big\rangle_{\rho\rho}$, it has a relatively narrow width of approximately 22 MeV. The prominent decay channels are $\Xi_c\pi$, $\Sigma_c^* K$ and $\Lambda_c\bar{K}$, with the branching ratios of
\[
\frac{\Gamma\big[\Xi_c'1D\big|J^P=5/2^+,3\big\rangle_{\rho\rho} \to \Xi_c\pi\big]}{\Gamma_{\text{Total}}} \simeq 17\%,
\]
\[
\frac{\Gamma\big[\Xi_c'1D\big|J^P=5/2^+,3\big\rangle_{\rho\rho} \to \Sigma_c^* K\big]}{\Gamma_{\text{Total}}} \simeq 21\%
\]
and
\[
\frac{\Gamma\big[\Xi_c'1D\big|J^P=5/2^+,3\big\rangle_{\rho\rho} \to \Lambda_c\bar{K}\big]}{\Gamma_{\text{Total}}} \simeq 15\%.
\]
Hence, The state $\Xi_c'1D\big|J^P=5/2^+,3\big\rangle_{\rho\rho}$ is very promising for experimental observation in decay channels $\Xi_c\pi$, $\Lambda_c\pi K$ and $\Lambda_c\bar{K}$.

The state $\Xi_c'1D\big|J^P=7/2^+,3\big\rangle_{\rho\rho}$ is a narrow state, with a width of approximately 14 MeV. The dominant decay modes are $\Xi_c'^*\pi$ and $\Sigma_c^* K$, and the corresponding branching ratios are
\[
\frac{\Gamma\big[\Xi_c'1D\big|J^P=7/2^+,3\big\rangle_{\rho\rho} \to \Xi_c'^*\pi\big]}{\Gamma_{\text{Total}}} \simeq 21\%
\]
and
\[
\frac{\Gamma\big[\Xi_c'1D\big|J^P=7/2^+,3\big\rangle_{\rho\rho} \to \Sigma_c^* K\big]}{\Gamma_{\text{Total}}} \simeq 42\%,
\]
respectively. Accordingly, collaborations can search for this state in the $\Xi_c\pi\pi$ and $\Lambda_c\pi K$ channels through the processes
$\Xi_c'1D\big|J^P=7/2^+,3\big\rangle_{\rho\rho}\to \Xi_c'^*\pi \to \Xi_c\pi\pi$ and
$\Xi_c'1D\big|J^P=7/2^+,3\big\rangle_{\rho\rho}\to \Sigma_c^* K \to \Lambda_c\pi K$, respectively.

\section{Summary}
In the present work, we systematically study the mass spectra of the $1S$-, $1P$- and $1D$-waves $\Xi_c$ and $\Xi_c'$ baryons within the potential quark model in the $j-j$ coupling scheme. Then, based on the calculated mass spectrum and wave functions, we explore the strong decay properties of the $1P$- and $1D$-wave excited states within the QPC model. We provide theoretical interpretations for the currently controversial resonances and make predictions for some "missing" resonances, which we hope will serve as a useful theoretical reference for future experimental searches. The main conclusions are summarized as follows.

For the two $1P$-wave $\lambda$-mode $\Xi_c$ baryons $\Xi_c1P\big|J^P=1/2^-,1\big\rangle_{\lambda}$ and $\Xi_c1P\big|J^P=3/2^-,1\big\rangle_{\lambda}$, their calculated masses and strong decay properties are in good agreement with those of the observed states $\Xi_c(2790)$ and $\Xi_c(2815)$, respectively. We therefore naturally interpret the observed states $\Xi_c(2790)$ and $\Xi_c(2815)$ as $\Xi_c1P\big|J^P=1/2^-,1\big\rangle_{\lambda}$ and $\Xi_c1P\big|J^P=3/2^-,1\big\rangle_{\lambda}$, respectively.

For the $1P$-wave $\lambda$-mode $\Xi_c'$ baryons, the predicted masses of $\Xi_c'\big|J^P=\tfrac12^-,0\big\rangle_{\lambda}$ and $\Xi_c'\big|J^P=\tfrac12^-,1\big\rangle_{\lambda}$ are $2882$ MeV and $2867$ MeV, respectively. $\Xi_c'1P\big|J^P=1/2^-,0\big\rangle_\lambda$ is likely to be a moderate state with a total decay width about $\Gamma\sim64$ MeV, and dominantly decays into the final states $\Xi_c\pi$ and $\Lambda_c\bar{K}$.
The total decay width of $\Xi_c'1P\big|J^P=1/2^-,1\big\rangle_\lambda$ is slightly narrower, around $\Gamma\sim25$ MeV, and is almost entirely saturated by the decay channel $\Xi_c'\pi$.  The $\Xi_c'1P\big|J^P=3/2^-,1\big\rangle_\lambda$ state could be a candidate for the observed state $\Xi_c(2923)$, as their masses and decay properties are in general agreement. $\Xi_c'1P\big|J^P=3/2^-,2\big\rangle_{\lambda}$ has a predicted mass of $2948$ MeV and a total decay width of approximately $13$ MeV, with $\Xi_c\pi$ and $\Lambda_c\bar{K}$ as its main decay modes. Both its mass and decay behaviors are consistent with those of $\Xi_c(2939)$. $\Xi_c'\big|J^P=\tfrac52^-,2\big\rangle_{\lambda}$ is predicted at $2969$ MeV with a total decay width around $12$ MeV. Its dominant decay channels are also $\Xi_c\pi$ and $\Lambda_c\bar{K}$, making it a plausible candidate for $\Xi_c(2965)$.

For the $1P$-wave $\rho$-mode $\Xi_c$ states, the predicted masses of $\Xi_c1P\big|J^P=1/2^-,0\big\rangle_{\rho}$ and $\Xi_c1P\big|J^P=1/2^-,1\big\rangle_{\rho}$ are $2834$ MeV and $2861$ MeV, respectively. Their decay properties are also quite similar: both have moderate total widths of $\Gamma \sim (30\text{--}40)$ MeV and decay dominantly into $\Xi_c\pi$, $\Xi_c'\pi$, and $\Lambda_c\bar{K}$. If we consider the heavy quark symmetry being not strictly true, then depending on the degree of symmetry breaking, $\Xi_c1P\big|J^P=1/2^-,0\big\rangle_{\rho}$ and $\Xi_c1P\big|J^P=1/2^-,1\big\rangle_{\rho}$ could both be candidates for the observed state $\Xi_c(2882)$. If our theoretical interpretation is correct, decay channels $\Xi_c\pi$ and $\Xi_c'\pi$ would also be promising channels for detecting state $\Xi_c(2882)$. For the other three $1P$-wave states, $\Xi_c1P\big|J^P=3/2^-,1\big\rangle_{\rho}$ has a predicted mass of $3005$ MeV and a total decay width of $\Gamma\sim 35$ MeV, with the $\Xi_c'^*\pi$ mode dominating its decays. The state $\Xi_c1P\big|J^P=3/2^-,2\big\rangle_{\rho}$ is predicted at $3002$ MeV with $\Gamma\sim 16$ MeV, and its prominent decay channels are $\Xi_c\pi$ and $\Lambda_c\bar{K}$. $\Xi_c1P\big|J^P=5/2^-,2\big\rangle_{\rho}$ lies at $3002$ MeV with a moderate width $\Gamma\sim 46$ MeV, and $\Xi_c\pi$ and $\Lambda_c\bar{K}$ are also among its major decay modes. These states are highly likely to be observed experimentally in their respective dominant decay channels.

For the $1P$-wave $\rho$-mode $\Xi_c'$ states, the predicted masses of $\Xi_c'1P\big|J^P=1/2^-,1\big\rangle_{\rho}$ and $\Xi_c'1P\big|J^P=3/2^-,1\big\rangle_{\rho}$ are $2946$ MeV and $2976$ MeV, respectively. Both states have total decay widths of approximately
 $\Gamma\sim 20$ MeV. The dominant decay mode of $\Xi_c'1P\big|J^P=1/2^-,1\big\rangle_{\rho}$ is $\Xi_c'\pi$, while the dominant decay mode of $\Xi_c'1P\big|J^P=3/2^-,1\big\rangle_{\rho}$ is $\Xi_c'^*\pi$. Therefore, experimental searches for these two states can be performed in the  $\Xi_c'\pi$ or $\Xi_c\pi\pi$ decay final states.

 For the $1D$-wave $\lambda$-mode $\Xi_c$ state $\Xi_c1D\big|J^P=3/2^+,2\big\rangle_{\lambda\lambda}$, our predicted mass is $3051$ MeV with a total decay width of $6.99$ MeV. Its dominant decay channels are $\Xi_c'\pi$ and $\Sigma_c K$. Considering the properties of the observed state $\Xi_c(3055)$,  $\Xi_c1D\big|J^P=3/2^+,2\big\rangle_{\lambda\lambda}$ is an ideal candidate for $\Xi_c(3055)$.  Meanwhile, if the state $\Xi_c(3055)$ is indeed interpretable as the quark configuration $\Xi_c1D\big|J^P=3/2^+,2\big\rangle_{\lambda\lambda}$, then, apart from the decay channel $\Sigma_cK$, the decay channel $\Xi_c'\pi$ can also be used as an alternative decay channel for the experimental study of state  $\Xi_c(3055)$.
 The state $\Xi_c1D\big|J^P=5/2^+,2\big\rangle_{\lambda\lambda}$ has a theoretical mass of $3054$ MeV and a total decay width of $7.33$ MeV, with the decay modes $\Xi_c'^*\pi$ and $\Sigma_c^* K$ dominating its decays. These properties are consistent with those of the observed state $\Xi_c(3080)$. Therefore,
$\Xi_c(3080)$ can be assigned as the quantum state $\Xi_c1D\big|J^P=5/2^+,2\big\rangle_{\lambda\lambda}$.

For the $1D$-wave $\lambda$-mode $\Xi_c'$ states, their masses lie in the range $(3100$–$3200)$ MeV. On the basis of their decay properties, we regard the two states $\Xi_c'1D\big|J^P=1/2^+,1\big\rangle_{\lambda\lambda}$ and $\Xi_c'1D\big|J^P=5/2^+,3\big\rangle_{\lambda\lambda}$ as potential candidates for $\Xi_c(3123)$, with the latter being more favorable. For the other four $1D$-wave $\lambda$-mode $\Xi_c'$ states, their decay widths are generally not broad, and these states are likely to be observed in their dominant decay channels, especially the channels with a $P$-wave state and a pion in the final state.

For the $1D$-wave $\rho$-mode $\Xi_c$ states, the predicted masses of $\Xi_c1D\big|J^P=3/2^+,2\big\rangle_{\rho\rho}$ and $\Xi_c1D\big|J^P=5/2^+,2\big\rangle_{\rho\rho}$ are $3249$ MeV and $3288$ MeV, respectively. The total decay width of $\Xi_c1D\big|J^P=3/2^+,2\big\rangle_{\rho\rho}$ is roughly $31$ MeV, with $\Sigma_c K$, $\Sigma_c^* K$, and $\Xi_c\big|J^P=\tfrac32^-,2\big\rangle_{\rho}$ dominating its decays. The state $\Xi_c1D\big|J^P=5/2^+,2\big\rangle_{\rho\rho}$ has a total decay width of about $50$ MeV, with dominant decay channels including $\Xi_c'^*\pi$, $\Sigma_c K$, $\Sigma_c^* K$, and $\Xi_c\big|J^P=\tfrac52^-,2\big\rangle_{\rho}$.

For the $1D$-wave $\rho$-mode $\Xi_c'$ states, all states fall in the mass window $3200$–$3300$ MeV, except for $\Xi_c'1D\big|J^P=1/2^+,1\big\rangle_{\rho\rho}$, which has a predicted mass of $3055$ MeV. $\Xi_c'1D\big|J^P=1/2^+,1\big\rangle_{\rho\rho}$ is probably a very narrow state, with a width below 1 MeV, while
 $\Xi_c'1D\big|J^P=3/2^+,1\big\rangle_{\rho\rho}$ and $\Xi_c'1D\big|J^P=3/2^+,2\big\rangle_{\rho\rho}$ possess moderate decay widths of several tens of MeV, which makes them relatively difficult to observe experimentally.
$\Xi_c'1D\big|J^P=5/2^+,2\big\rangle_{\rho\rho}$ is very likely to be a moderately broad state, with a width of about 40 MeV, and decays predominantly via the $\Sigma_c^*K$ channel. The state $\Xi_c'1D\big|J^P=5/2^+,3\big\rangle_{\rho\rho}$ has a total decay width of approximately $22$ MeV, and it decays predominantly to $\Xi_c\pi$, $\Sigma_c^* K$, $\Lambda_c\bar{K}$, and $\Xi_c'1P\big|J^P=3/2^-,1\big\rangle_{\rho}\pi$. For $\Xi_c'1D\big|J^P=7/2^+,3\big\rangle_{\rho\rho}$, the total decay width is around $14$ MeV; its dominant decay modes are $\Xi_c'^*\pi$ and $\Sigma_c^* K$, with possible cascades to the final states $\Xi_c\pi\pi$ and $\Lambda_c\pi K$.

\section*{Acknowledgements }

This work is supported by the National Natural Science Foundation of China under Grants No.12005013 and Fundamental Research Funds for the Central Universities (No. FRF-BRB-25-006).

\bibliographystyle{unsrt}
\bibliography{Xic20261}



	\end{document}